\input harvmac
\def\a{\alpha}    \def\b{\beta}              
\def\d{\delta}
    
\def\e{\varepsilon} 
\def\ep{\epsilon}

\def\g{\gamma}    \def\G{\Gamma}      \def\k{\kappa}     
\def\l{\lambda}
\def\L{\Lambda}   \def\m{\mu}         \def\n{\nu}        
\def\r{\rho}
  \def\o{\omega}      \def\O{\Omega}     
\def\p{\psi}
      \def\s{\sigma}      \def\S{\Sigma}     
\def\th{\theta}
\def\t{\tau}

\def\CA{{\cal A}}
\def\CB{{\cal B}}

\def\CG{{\cal G}}
\def\CM{{\cal M}}
\def\CN{{\cal N}}

\def\CW{{\cal W}}

%
\font\teneufm=eufm10
\font\seveneufm=eufm7
\font\fiveeufm=eufm5
\newfam\eufmfam
\textfont\eufmfam=\teneufm
\scriptfont\eufmfam=\seveneufm
\scriptscriptfont\eufmfam=\fiveeufm
\def\eufm#1{{\fam\eufmfam\relax#1}}

\font\teneusm=eusm10
\font\seveneusm=eusm7
\font\fiveeusm=eusm5
\newfam\eusmfam
\textfont\eusmfam=\teneusm
\scriptfont\eusmfam=\seveneusm
\scriptscriptfont\eusmfam=\fiveeusm

\font\tenmsx=msam10
\font\sevenmsx=msam7
\font\fivemsx=msam5
\font\tenmsy=msbm10
\font\sevenmsy=msbm7
\font\fivemsy=msbm5
\newfam\msafam
\newfam\msbfam
\textfont\msafam=\tenmsx  \scriptfont\msafam=\sevenmsx
  \scriptscriptfont\msafam=\fivemsx
\textfont\msbfam=\tenmsy  \scriptfont\msbfam=\sevenmsy
  \scriptscriptfont\msbfam=\fivemsy

\def\msbm#1{{\fam\msbfam\relax#1}}

\def\rd{\partial}

\def\darr#1{\raise1.5ex\hbox{$\leftrightarrow$}\mkern-16.5mu #1}

\def\Fr#1#2{{#1\over#2}}
\def\tr{\hbox{Tr}\,}

\def\roughly#1{\raise.3ex\hbox{$#1$\kern-.75em
\lower1ex\hbox{$\sim$}}}

\def\pr{\prime}

\def\cmp#1#2#3{Comm.\ Math.\ Phys.\ {{\bf #1}} {(#2)} {#3}}
\def\pl#1#2#3{Phys.\ Lett.\ {{\bf #1}} {(#2)} {#3}}
\def\np#1#2#3{Nucl.\ Phys.\ {{\bf #1}} {(#2)} {#3}}

\def\ijmp#1#2#3{Int.\ J.\ Mod.\ Phys.\ {{\bf #1}} 
{(#2)} {#3}}

\def\top#1#2#3{Topology {{\bf #1}} {(#2)} {#3}}

\font\tenbifull=cmmib10 
\font\tenbimed=cmmib10 scaled 800
\font\tenbismall=cmmib10 scaled 666
\textfont9=\tenbifull \scriptfont9=\tenbimed
\scriptscriptfont9=\tenbismall

\def\Bg{{\fam=9{\mathchar"7111} }}

\def\Bu{{\fam=9{\mathchar"711F } }}
\def\Bv{{\fam=9{\mathchar"7120 } }}

\def\bs{{\bf s}}
\def\bbs{{\bf\bar s}}
\def\Dp{\rd_{\!A}}
\def\Dpp{\bar\rd_{\!A}}
\magnification=1200
\def\lin#1{\noindent {$\underline{\hbox{\it #1}}$} }
\lref\VW{
C.~Vafa and E.~Witten,
\np{B 431}{1994}{3}.
}
\lref\SWa{
N.~Seiberg and E.~Witten,
\np{B 426}{1994}{19}.
}
\lref\SWb{
N.~Seiberg and E.~Witten,
\np{B 431}{1994}{484}\semi
R.~Donagi and E.~Witten,
\np{B 460}{1996}{299}. 
}
\lref\WittenA{
E.~Witten,
\cmp{117}{1988}{386};
\ijmp{A6}{1991}{2273}
}
\lref\WittenB{
E.~Witten,
J.~Math.~Phys.~{\bf 35} (1994) 5101.
}
\lref\WittenC{
E.~Witten,
Math.~Res.~Lett.~{\bf 1} (1994) 769.
}
\lref\DK{
S.K.~Donaldson and P.B.~Kronheimer,
{\it The geometry of four-manifolds},
Oxford Mathematical Monographs, Clarendon Press, Oxford,
1990
}
\lref\HYM{
J.-S.~Park,
\cmp{163}{1994}{113};
\np{B423}{1994}{559}
\semi
S.~Hyun and J.-S.~Park,
J.~Geom.~Phys.~{\bf 20} (1996) 31.
}
\lref\Donaldson{
S.K.~Donaldson,
\top{29}{1990}{257}
}
\lref\DM{
R. Dijkgraaf and G. Moore,
\cmp{185}{1997}{411}.
}
\lref\Taubes{
C.~H.~Taubes,
Math.~Res.~Lett.~{\bf 1} (1995) 221
}
\lref\BT{
M.~Blau and G.~Thompson,
\cmp{152}{1993}{41};
\ijmp{A 8}{1993}{573}.
}
\lref\CMR{
S.~Cordes, G.~Moore and S.~Ramgoolam,
\cmp{185}{1997}{543};
Nucl.~Phys.~Proc.~Suppl.~41(1995)184.
}
\lref\DUAL{
C.~Montonen and D.~Olive,
\pl{B72}{1977}{117}\semi
E.~Witten and D.~Olive,
\pl{B78}{1978}{97}\semi
H.~Osborn,
\pl{B83}{1979}{321}
}
\lref\Yamron{
J.~Yamron, \pl{B213}{1988}{353}
}
\lref\Monads{
J.-S.~Park,
\np{B 493}{1997}{198}.
}
\lref\WittenF{
E.~ Witten, 
J. Geom. Phys. {\bf G9} (1992) 303.
}
\lref\MW{
G.~Moore and E.~Witten,
{\tt hep-th/9709193}.
}

\lref\LLA{ J.~M.~F.~Labastida and C.~Lozano, 
\np{B 502}{1997}{741}.
}

\lref\LL{ J.~M.~F.~Labastida and C.~Lozano,
{\tt hep-th/9711132}.
}

\lref\blau{
M.~Blau and G.~Thompson,
\np{B 492}{1997}{545}
}

\lref\Wittenmirror{
E.~Witten,
{\tt hep-th/9112056}.
}
 
\lref\GLSM{
E.~Witten,
Nucl.Phys. B403 (1993) 159.
}

\lref\Wittengr{
E.~Witten,
{\tt hep-th/9312104}.
}

\lref\juanLN{
J.M.~Maldacena
{\tt hep-th/9711200}
}

\font\Titlerm=cmr10 scaled\magstep3
\nopagenumbers
\rightline{ITFA-97-09, hep-th/9801066
}

\vskip .7in

\centerline{\fam0\Titlerm 
N=4 Supersymmetric Yang-Mills Theory
}
\vskip 0.2in
\centerline{\fam0\Titlerm 
on a K\"{a}hler Surface}

\tenpoint\vskip .4in\pageno=0

\centerline{Robbert Dijkgraaf}
\vskip .1in
\centerline{\it
Department of Mathematics}
\centerline{\it
University of Amsterdam, 1018 TV Amsterdam}
\vskip .2in
\centerline{ 
Jae-Suk Park and Bernd J.~Schroers}
\vskip .1in
\centerline{\it Institute for Theoretical Physics}
\centerline{\it University of Amsterdam, 1018 XE Amsterdam}

\vskip .4in

\noindent
\abstractfont

We study $N=4$ supersymmetric Yang-Mills theory on a
K\"{a}hler manifold with $b_2^+\geq 3$.
Adding suitable perturbations
we show that the partition function of the $N=4$ theory 
is the sum of contributions from two branches:
(i) instantons,  (ii) a special class of 
Seiberg-Witten monopoles.  We determine the partition 
function for  the theories with  gauge group $SU(2)$ and $SO(3)$,
using  $S$-duality. This leads us to a formula
for  the Euler characteristic of the moduli space
of instantons.

\vskip 0.3in

\Date{December 1997; Revised by October 1998}


\newskip\normalparskip
\normalparskip = 5pt plus 1.2pt minus .6pt
\parskip = \normalparskip

\baselineskip=14pt plus 1.2pt minus .6pt


\newsec{Introduction}

In the paper \VW\ Vafa and Witten presented strong evidence
for  $S$-duality
of $N=4$ super-Yang-Mills theory \DUAL.
They used  a topological,  twisted version of the $N=4$ theory
\WittenA\Yamron\ and were  able to
determine  the partition function  of $N=4$
super-Yang-Mills theory on certain K\"ahler manifolds.
In particular they  identified the partition
function with the Euler characteristic of the
moduli space of instantons, provided certain vanishing theorems hold.

This paper is an elaboration and generalisation of the work of Vafa and
Witten. We want to determine the partition function for a general
compact K\"{a}hler surface $X$ with $b_2^+ \geq 3$. The twisted version
of $N=4$ super-Yang-Mills theory studied by Vafa and Witten is an
important example of a balanced topological field theory as defined in
\DM\ (see also \BT\CMR\blau\LLA). These topological field theories carry two
topological supercharges. On a K\"ahler manifold the number of
topological charges is extended to four, and these charges may be
interpreted as the differentials of a Dolbeault version of balanced
$\CG$-equivariant cohomology \HYM, where $\CG$ is the group of gauge
transformations. Our computation of the partition function involves a a
series of perturbations which break the supersymmetry down to $N=2$ and
$N=1$ (topological) supersymmetry.

The perturbation down to $N=2$ is achieved by adding a  bare mass term 
for the
hypermultiplet. Geometrically, this term may be viewed as the 
equivariant momentum map of a  $\CG\times S^1$-action 
on the hypermultiplet.
As a result of its inclusion in the  action, 
the path integral is localised on the fixed point set of
the $\CG\times S^1$-action, which consists of two branches: (i) the
moduli space of anti-self-dual connections, (ii) the moduli space of a
certain class of Seiberg-Witten monopoles \WittenC. Perturbing further
down to $N=1$, following \WittenB, leads to the factorization of the
Seiberg-Witten classes contributing to branch (ii).

Specializing to gauge groups $SU(2)$ and $SO(3)$ we propose a formula
for the branch (ii) contribution on a general K\"ahler manifold with
$b_2^+ \geq 3$. Then $S$-duality of $N=4$ super-Yang-Mills theory
enables us to determine the entire partition function. As a corollary we
obtain a formula for the Euler characteristic of the moduli space of
instantons (branch (i)). We consider the pure $N=2$ limit and
obtain the essential part of Witten's formula for Donaldson invariants
\WittenC. In the final section we add brief comments on relation with
a $\CN_{ws}=(2,2)$ gauged linear sigma model and its applications. 

\newsec{Prelude}

Here we briefly review the notion of a balanced topological Yang-Mills
theory \DM, which is a twisted $N=4$ theory \VW. The topological field
theory computing the Euler characteristic of certain moduli spaces was
studied in \BT\ and \CMR. (See \blau\ for the precise relation of the
earlier work \BT\ with \DM.) We also recall salient features of
topological Yang-Mills theory \WittenA\ on a K\"{a}hler manifold
\HYM\WittenF.

\subsec{ Balanced Topological Yang-Mills Theory} 

Let $X$ be a compact Riemann four manifold and
$E$ be a $SU(n)$-bundle over $X$. The bundle $E$
is classified by the instanton number
\eqn\inst{
k =\Fr{1}{8\pi^2}\int_X \tr F\wedge F,
}
where $\tr$ is the trace in the fundamental representation
of $SU(n)$ and $F\in \O^2_X(\eufm{g}_E)$ is the adjoint
valued curvature $2$-form on $M$. We denote 
the group of gauge transformation by $\CG$, i.e.
elements of $ \CG$  are maps  $g:X\rightarrow SU(N)$.

We introduce two global supercharges 
$Q_\pm$ carrying an additive quantum number (ghost number)
$U=\pm 1$. They satisfy the following commutation
relations:
\eqn\balance{
Q_+^2 =\d_{\phi_{++}},\qquad
\{Q_+, Q_-\}= \d_{\phi_{+-}},\qquad
Q_-^2 =\d_{\phi_{--}}.
}
where $\d_{\phi}$ denotes the gauge transformation generated by an
adjoint scalar $\phi\in\O^0_X(\eufm{g}_E)$. The charges $Q_\pm$ can be
interpreted physically as the twisted supercharges of $N=4$
super-Yang-Mills theory \VW\ or mathematically as the differentials of
balanced $\CG$-equivariant cohomology \DM.

To have a complete representation of this algebra
one needs to impose a consistency condition (Bianchi identity) 
which is  conveniently summarised in the following quintet of fields
\eqn\ymvi{
\matrix{
U=+2\cr
{}\cr
U=+1\cr
{}\cr
U=0\cr
{}\cr
U=-1\cr
{}\cr
U=-2\cr
}\qquad
\qquad {\rm consistency}\quad
\matrix{
\phi_{++}&          &      \cr
         &\searrow  &      \cr
         &          &\Bg_+\cr
         &\nearrow  &      \cr
\phi_{+-}&          &      \cr
         &\searrow  &      \cr
         &          &\Bg_-\cr
         &\nearrow  &      \cr
\phi_{--}&          &      \cr
}.
}
Here $\nearrow$ and $\searrow$ represent the action of $Q_+$ and $Q_-$
respectively. The diagram then expresses the requirement that acting
with $Q_+$ and $Q_-$ on the triplet $(\phi_{++},\phi_{+-},\phi_{--})$
one generates in the indicated manner a space which is spanned by the
doublet $(\Bg_+, \Bg_-)$. Since our global supercharges are scalars we
have $\Bg_\pm\in \O^0_X(\eufm{g}_E)$. We will sometimes use the notation
$\phi_{++}=\phi$, $\phi_{+-} =C$ and $\phi_{--} =\bar\phi$.

Next we have to choose the basic field of our theory.
Here we take a   connection $1$-form  $A$. 
Acting
with the supercharges will then generate 
three further   fields  $\Bv_{\pm},H\in \O^1_X(\eufm{g}_E)$.
This leads to the basic quartet of any balanced topological
field theory.
To define the theory we want to study here 
we also need 
a self-dual adjoint-valued  two-form  $B$, which leads to 
another quartet. These two multiplets are 
summarised in the following picture:
\eqn\matter{\matrix{
U=+1\cr
{}\cr
U=0\cr
{}\cr
U=-1\cr
}\qquad
{\rm fields}\quad
\matrix{    &  &   \Bv_{+}  &  &  \cr
    & \nearrow &  &  \searrow & \cr
A &  & & & H\cr
 & \searrow & & \nearrow & \cr
 & & \Bv_{-}& & \cr},
\qquad
\qquad {\rm equations}\quad
\matrix{    &  &   \Bu_{+}    &  &  \cr
    & \nearrow &  &  \searrow & \cr
B  &  & & & H^+   \cr
 & \searrow & & \nearrow & \cr
 & &  \Bu_{-} & & \cr},
}
where
$\Bu_\pm,H^+ \in \O^{2+}_X(\eufm{g}_E)$.
A balanced topological field theory 
 is uniquely determined by its field content
and the  balanced algebra, and moreover has
 the pleasant property that there is no $U$-number
anomaly: any fermionic zero-mode
has a partner with the opposite $U$-number.

The action functional can be written as
\eqn\aaf{
S = \Fr{1}{2}(Q_+Q_- -Q_-Q_+)\left(\CF_0 +\CF_1\right)
}
where
\eqn\aafc{\eqalign{
&\CF_0 = -\Fr{4}{e^2}\int_X\!\!\tr\biggl(
B^{\m\n}\left(F^+_{\m\n} + \Fr{1}{12} [B_{\m\r}, B_{\n\s}]g^{\r\s}
\right)
\biggr), 
\cr
&\CF_1 = -\Fr{4}{e^2}\int_X\!\!\tr\biggl(
\Bu_+\!\wedge \Bu_- 
-\Bv_+\!\wedge*\Bv_-
+\Bg_+\!*\Bg_-
\biggr),
}
}
The bosonic part  of the action functional has the following explicit
form
\eqn\actiona{\eqalign{
S_{bose}=
\Fr{4}{e^2}&\int_X \tr\biggl(
H^+\!\wedge\! H^+ 
- H^{\m\n +}\left(F^+_{\m\n} + \Fr{1}{4}[B_{\m\r},B_{\n\s}]g^{\r\s}\right)
+ H\!\wedge\! * H 
- B\!\wedge\!  d_{\!A}^+H 
\phantom{\biggr)}
\cr
& \phantom{\biggl(}
-\Fr{1}{4}d_{\!A} C\!\wedge\! * d_{\!A} C
- \Fr{1}{4}[C,B]^2
+[\phi,\bar\phi]^2
+d_A\phi* d_A\bar\phi
+[\phi,B]\wedge [\bar\phi,B]
\phantom{\biggr)}
\cr
&\phantom{\biggl(}
+[\phi,C]*[\bar\phi,C]
\biggr),
}
}
where $d_{\!A}^+$ denotes the projection of 
$d_{\!A}$ to  the self-dual part.
We integrate out $H^+$ and $H$ 
using the algebraic equation of motions
\eqn\alge{
H_{\m\n}^+ =\Fr{1}{2}\left( F_{\m\n}^+ 
+ \Fr{1}{4}[B_{\m\r},B_{\n\s}]g^{\r\s}\right),\qquad
H =\Fr{1}{2} d_{\!A}^{*}B,
}
and a  Weitzenb\"{o}ck formula. We get
\eqn\action{\eqalign{
S_{bose} =-
\Fr{1}{e^2}&\int_X \tr\biggl(
F^+\!\wedge\! F^+ 
+\Fr{1}{4}d_{\!A} B\!\wedge\!* d_{\!A}B
+d_{\!A} C\!\wedge\! * d_{\!A} C
+\Fr{1}{16}[B_{\m\r},B_{\n\r}][B_{\m\s}, B_{\n\s}]
\cr
&
+ [C,B^+]^2
+\Fr{1}{4}B_{\m\n}\left(
\Fr{1}{6}(g_{\m\rho}g_{\n\s} -g_{\m\s}g_{\n\rho})R +
W^+_{\m\n\rho\s}\right)B_{\rho\s} +4[\phi,\bar\phi]^2
\cr
&
-4d_A\phi* d_A\bar\phi
-4[\phi,B]\wedge[\bar\phi, B]
-4[C,\phi]*[C,\bar\phi]
\biggr),
}
}
where $R$ is the scalar curvature of $X$ and $W^+$ is the self-dual
part of the Weyl tensor.
The resulting action is invariant under $Q_\pm$ if we 
modify the transformation laws of $\Bu_{\pm}$ according to
the replacement \alge.
Due to the fixed point theorem of Witten for global supersymmetry
the path integral is localized on the  fixed point
locus of $Q_\pm$. This locus is given by the following equations:
\eqn\fixed{
\eqalign{
&F^+_{\m\n} +\Fr{1}{4}[B_{\m\r},B_{\n\s}]g^{\r\s}=0, ,\cr
&d_{\!A}^{*}B=0,\cr
}
\qquad
\eqalign{
&[C,B]=[\phi,\bar\phi]=[C,\phi_{\pm\pm}]=[\phi_{\pm\pm},B]=0,\cr
&d_{\!A} C=d_{\!A}\phi_{\pm\pm}=0.\cr
}
}
These  equations  are equivalent to 
the  equations in section 2.4  of  \VW\  and we call them 
the Vafa-Witten equations.
The equations for fermionic zero-modes are just the linearization
of the fixed point equation \fixed\ and the condition that they
are orthogonal to gauge orbits.
Due to the balanced structure each fermionic zero-mode
has  a partner with the opposite $U$-number.
Thus there is no ghost-number anomaly and
the partition function is well-defined.

\subsec{Topological Yang-Mills Theory}

Both $Q_+$ and $Q_-$ individually satisfy the $N_T=1$
supersymmetry algebra; mathematically both 
may thus be thought of as differentials of $\CG$-equivariant cohomology.
Concentrating on $Q_+$  we obtain
 the algebra for the original topological Yang-Mills theory \WittenA.
\eqn\qru{
Q_+^2=\d_{\phi_{++}},\qquad
A\longrightarrow \Bv_+,\quad
\phi_{--}\longrightarrow \Bg_-,\quad
\Bu_- \longrightarrow H^+,
}
Now we no longer have a balanced structure and expect U-number 
anomalies.
This model has well-known non-trivial observables
representing Donaldson's map \Donaldson
\eqn\ake{\eqalign{
&\CO_0= \Fr{1}{8\pi^2}\int_X\tr(F\!\wedge\! F),\cr
&\CO_2= \Fr{1}{4\pi^2}
\int_{\g_2}\tr(\phi_{++} F+\Bv_+\!\wedge\!\Bv_+),\cr
&\CO_4= \Fr{1}{8\pi^2}\tr(\phi_{++}^2),\cr
}\qquad
\eqalign{
&\CO_1=\Fr{1}{4\pi^2}\int_{\g_3}\tr(\Bv_+\!\wedge\! F),\cr
&\CO_3=\Fr{1}{4\pi^2}\int_{\g_1}\tr(\phi_{++}\Bv_+),\cr
}
}
where $\g_i \in H_i(X)$ and the subscript denotes the $U$-number.
The expectation values of the observables define
Donaldson invariants.

We shall now see that one can obtain topological Yang-Mills 
theory from balanced topological Yang-Mills 
theory  by introducing
mass terms for $\Bv_-,\phi_{+-}, \Bg_+, B^+, \Bu_+$,
provided the underlying four-manifold is  K\"{a}hler.

\subsec{Topological Yang-Mills theory on a K\"ahler manifold}

Now let $X$ be a K\"{a}hler surface with the K\"{a}hler form $\o$.
For a $SU(n)$ bundle $E$ over $X$ we have the space $\CA$
of all connections, which is an affine space. A 
tangent vector on $\CA$ can be
represented by an element of
$\O^1_X(\eufm{g}_E)$.
Picking a complex structure $I$ in $X$ we can introduce
a complex structure on $\CA$ by the decomposition 
$T\CA = T^{1,0}\CA\oplus T^{0,1}\CA$ and the identification
$\d A^{0,1} \in   T^{1,0}\CA$ where 
$\d A^{0,1} \in \O^{0,1}_X(\eufm{g}_E)$. We also have  a 
K\"{a}hler structure on $\CA$ with the K\"{a}hler form
\eqn\fdas{
\tilde \o =\Fr{1}{4\pi^2}\int_X\tr(\d A^{1,0}\wedge \d A^{0,1})\wedge\o.
}
Any self-dual two-form $\a^+$
can be decomposed as
\eqn\aag{
\a^+ = \a^{2,0} + \a^0\o + \a^{0,2},
}
where $\a^{0,2}\in\O^{0,2}_X$ and $\a^0 \in \O^0(X)$.
In particular, an anti-self-dual connection $A$ satisfies
\eqn\sfsa{
F^{2,0}_A = 0, \qquad F_A\wedge \o=0,\qquad F_A^{0,2}=0.
}
We denote the  moduli space of anti-self-dual connections by
$\CM$.

Since a K\"{a}hler surface has $U(2)$ holonomy,
the number $N_T$ of topological
supersymmetries is doubled.
Thus $N_T$ is identical to the number $N$ of physical
supersymmetries of the underlying super-Yang-Mills theory.
So we will use $N$ to denote  both the number of
topological and physical supersymmetries.
Then the  supersymmetry  generator $Q_+$ \qru\ 
can be  decomposed as $Q_+ = \bs_+ + \bbs_+$, with respect
to the complex structure on $\CA$, with
the commutation relation
\eqn\fdfd{
\bs_+^2 =0,\qquad \{\bs_+,\bbs_+\}= \d_{\phi_{++}},
\qquad \bbs_+^2=0.
}
The supercharges $\bs_+$ and $\bbs_+$  are the differentials of 
the Dolbeault version
of $\CG$-equivariant cohomology \HYM.
There is a   natural bi-grading $(p,q)$ such that 
the original ghost number $U$  is given by $U=(p+q)$ and
there is  a new ghost number $R$ given by $R=(p-q)$. 
The $N=2$ supercharges
$\bs_+$ and $\bbs_+$ are 
assigned the following degrees 
\eqn\derg{
\bs_+: (1,0),\qquad
\bbs_+: (0,1).
}
The connection one-form $A$ has the degree
$(0,0)$.
Indicating the action of  $\bs_+$ by $\nearrow$ and
the action of  $\bbs_+$ by $\searrow$,
the field  contents  and  $R$-number assignments  are
\eqn\tfiel{
\matrix{
R=+1\cr
{}\cr
R=0\cr
{}\cr
R=-1\cr
}
\quad
\matrix{    &  &   \p^{0,1}_{+}  \cr
    & \nearrow &  \cr
A   &  & \cr
 & \searrow \cr
 & & \bar\p_{+}^{1,0}\cr},
\quad
\phi_{++},
\quad
\matrix{    &  &   \eta_{-}    &  &  \cr
    & \nearrow &  &  \searrow & \cr
\phi_{--}  &  & & & H^0   \cr
 & \searrow & & \nearrow & \cr
 & &  \bar\eta_{-} & & \cr},
\quad
\matrix{\bar\chi^{0,2}_{-}    &  &  \cr
&  \searrow & \cr
& & H^{2,0},H^{0,2}   \cr
& \nearrow & \cr
\chi^{2,0}_{-} & & \cr}.
}
Here $\p^{0,1}_+\in\O^{0,1}_X(\eufm{g}_E)$, 
$\bar\p^{1,0}_+\in\O^{1,0}_X(\eufm{g}_E)$,
$\eta_-,\bar\eta_-,H^0 \in\O^0_X(\eufm{g}_E)$,
$\chi^{2,0}_-,H^{2,0}\in\O^{2,0}_X(\eufm{g}_E)$,
and $\bar\chi^{0,2}_-,H^{0,2}\in\O^{0,2}_X(\eufm{g}_E)$.
This diagram  shows that the model is balanced
in terms of the $R$-number. The net $U$-number, however, 
is generally non-vanishing and
equals  the formal dimension of the moduli space of anti-self-dual
 connections. The action functional of the theory is defined
by
\eqn\actionnt{
\eqalign{
S_2=&-\Fr{\bs_+}{e^2}\!\int_X\tr\chi^{2,0}_-\wedge F^{0,2}
-\Fr{\bbs_+}{e^2}\!\int_X\tr \bar\chi^{0,2}_-\wedge F^{2,0}
\cr
&+\Fr{\bs_+\bbs_+}{e^2}\int_X\tr\biggl(2i\phi_{--}F\wedge\o 
+\chi^{2,0}_-\wedge\bar\chi^{0,2}_-
-2\eta_- *\bar\eta_-\biggr)
}
}

One can further  break half of the $N=2$ symmetry by introducing
mass for  the $N=1$ matter multiplet ($N=1$ chiral multiplet) \WittenB.
The required  perturbation  makes use of a 
holomorphic two-form $\o^{0,2}\in \O^{0,2}_X$ 
and was essential  in the development of 
Donaldson and Seiberg-Witten  theory.
One can also deform  topological Yang-Mills 
theory to  holomorphic Yang-Mills 
theory \HYM.

\newsec{$N=4$ Theory}

In this section we study $N=4$ super-Yang-Mills
theory on a K\"{a}hler surface $X$.

\subsec{$N=4$ Supersymmetry}

Naturally we can carry out the 
decomposition of $Q_+$ also for  $Q_-$.
Thus we write 
$Q_\pm = \bs_\pm + \bbs_\pm$
and have
\eqn\faaa{
\bs_\pm^2=0,\qquad\{\bs_\pm,\bbs_\pm\} 
= \d_{\phi_{\pm\pm}},
\qquad
\bbs_\pm^2=0.
}
We assign the degree $(p,q)$ to the super-charges
$\bs_\pm,\bbs_\pm$ by
\eqn\ghost{\eqalign{
&\bs_+ : (+1,0),\qquad \bbs_+ : (0,+1),\cr
&\bs_- : (0,-1),\qquad \bbs_- : (-1,0).\cr
}
}
We also introduce another bigraded
additive quantum number $(J_L,J_R)_\CR$, which we refer to
$U(1)_\CR$ charge.\foot{This terminology will be clarified in
a later section.} 
\eqn\rcharge{\eqalign{
&\bs_+ : (+1,0),\qquad \bbs_+ : (-1,0),\cr
&\bs_- : (0,-1),\qquad \bbs_- : (0,+1).\cr
}
}
Both the degree and $U(1)_\CR$ charge originate from
the unbroken part of $SO(5,1)_I$ symmetry of physical $N=4$
SYM theory after twisting on a K\"{a}hler manifold \VW.

The degree above represent the form degree in the
Dolbeault equivariant cohomology. The operators 
$\eufm{s}= \bs_+ \oplus\bs_-$ and $\bar\eufm{s} = \bbs_+ \oplus\bbs_-$
may be interpreted as the holomorphic
and the anti-holomorphic differentials 
 on iterated superspace \DM. Then, similarly to
\fdfd, we  have 
\eqn\sab{
\eufm{s}^2 = 0,\qquad \{\eufm{s},\bar\eufm{s}\}=\d_{\Phi},
\qquad \bar\eufm{s}^2=0,
}
where $\Phi$ is the gauge transformation generator.
According to our grading, we can decompose
$\Phi$ into components 
$\Phi = \phi_{++} \oplus \s\oplus\bar\s\oplus \phi_{--}$
with the degree $(1,1)\oplus (0,0)\oplus(0,0)\oplus(-1,-1)$
and  the $U(1)_\CR$ charge $(0,0)\oplus (1,1)\oplus (-1,-1)\oplus(0,0)$, 
respectively.
Now \sab\ reduces to 
\eqn\sac{
\eqalign{
\{\bs_+,\bbs_+\} = \d_{\phi_{++}},\cr
\{\bs_-,\bbs_-\} = \d_{\phi_{--}},\cr
}\qquad
\eqalign{
\{\bs_+,\bbs_-\}=\d_\s,\cr 
\{\bs_-,\bbs_+\} = \d_{\bar\s},\cr
}
\qquad
\eqalign{
\{\bs_+,\bs_-\}=0,\cr
\{\bbs_+,\bbs_-\}=0.\cr
}\qquad
\eqalign{
\bs_\pm^2=0,\cr
\bbs_\pm^2=0.
}
}

Note that 
\eqn\useles{
\{Q_+,Q_-\}=
\{\bs_+,\bbs_-\}+ \{\bs_-,\bbs_+\} = \d_{\phi_{+-}},
}
leading to $\s +\bar\s = \phi_{+-}$.
We decompose the  self-dual $2$-form $B\in \O^{2+}_X(\eufm{g}_E)$
as $B = B^{2,0} + B_0\o + B^{0,2}$. We combined the scalar
$B_0$ with
$\phi_{+-}$ into the complex scalar 
\eqn\yer{
\s\equiv \Fr{1}{2}\phi_{+-}+ \Fr{i}{2} B^0.
}

\subsec{The Fields and Action Functional} 

 \def\mapr{\!\smash{
    \mathop{\longrightarrow}\limits^{s_+}}\!}
\def\mapl{\!\smash{
    \mathop{\longleftarrow}\limits^{s_-}}\!}
\def\mapd{\Big\downarrow
  \rlap{$\vcenter{\hbox{$\scriptstyle \bar s_-$}}$}}
\def\mapu{\Big\uparrow
  \rlap{$\vcenter{\hbox{$\scriptstyle\bar s_+$}}$}}

It is convenient to represent the action of various supercharges
by  two-dimensional vectors as follows
\eqn\cgh{
\matrix{
 &\mapu &\cr
\mapl&  &\mapr\cr
 &\mapd&\cr
}
}
We  represent the various fields by points in the plane.
Then, the commutation relations \sac\ can be represented
by simple diagrammatic rules.

In order to  obtain  all the components
of the $N=4$ superfields
it is necessary and sufficient to introduce 
two basic fields, the connection $A$ and an adjoint
$(2,0)$-form $B^{2,0}$. 
We decompose the connection $1$-form $A$, 
$A = A^{1,0}+ A^{0,1}$ according to the complex structure
on $X$. Now we consider the space $\CA$ of all connections 
and the space of $\O_X^{2,0}(\eufm{g}_E)\oplus \O_X^{0,2}(\eufm{g}_E)$
all $B^{2,0}$ and $B^{0,2}$. We introduce a complex structure
on these space by declaring $A^{0,1}$ and $B^{2,0}$ as the
holomorphic coordinates. Then we have
\eqn\hoq{
\bar\eufm{s} A^{0,1}=0,\qquad \bar\eufm{s} B^{2,0}=0,
}
where $\bar\eufm{s} =\bbs_+\oplus \bbs_-$ as defined earlier.
From the above condition and the commutation relations \sac\ we build up  two 
holomorphic quartets  as follows;
\eqn\ffa{

\matrix{
\p^{0,1}_-&\mapl&A^{0,1}  &\mapr&\p^{0,1}_+\cr
&&&&\cr
\p^{0,1}_-&\mapr&H^{0,1} &\mapl&\p^{0,1}_+\cr
}\qquad\hbox{and}\qquad
\matrix{
\chi^{2,0}_-&\mapl&B^{2,0}   &\mapr&\chi^{2,0}_+\cr
&&&&\cr
\chi^{2,0}_-&\mapr&H^{2,0}  &\mapl&\chi^{2,0}_+, \cr
}
}
where $\p_\pm, H^{1,0}\in\O^{1,0}_X(\eufm{g}_E)$
and $\chi^{0,2}_\pm, H^{0,2}\in \O^{0,2}_X(\eufm{g}_E)$.
It follows that
\eqn\ahoq{
\eufm{s} A^{1,0}=0,\qquad \eufm{s} B^{0,2}=0,
}
which lead to two
anti-holomorphic quartets 
\eqn\ffb{

\matrix{
 &\bar\p^{1,0}_+&\cr
 &\mapu&\cr
 &A^{1,0}&\cr
&\mapd&\cr
 &\bar\p^{1,0}_-&,\cr
}
\qquad
\matrix{
 &\bar\p^{1,0}_+&\cr
 &\mapd&\cr
 &H^{1,0}&\cr
&\mapu&\cr
 &\bar\p^{1,0}_-  &,\cr
}\qquad
\hbox{and}
\qquad
\matrix{
 &\bar\chi^{0,2}_+&\cr
 &\mapu&\cr
 &B^{0,2}&\cr
&\mapd&\cr
 &\bar\chi^{0,2}_-  &,\cr
}
\qquad
\matrix{
 &\bar\chi^{0,2}_+&\cr
 &\mapd&\cr
 &H^{0,2}&\cr
&\mapu&\cr
 &\bar\chi^{0,2}_-  &.\cr
}
}
Imposing  the commutation relations \sac\ we obtain
the following  consistency nine-plet:
 \eqn\ffcccc{

\matrix{
 \bar\s       & \mapr &\eta_+& \mapl& \phi_{++}     \cr
  \mapd      &            &\mapd &          &\mapd \cr
\bar\eta_- &\mapr & H^0    &\mapl &\bar\eta_+ \cr
\mapu        &            &\mapu   &          &\mapu \cr
\phi_{--}&\mapr  & \eta_-      & \mapl  &\s          \cr
}
}
We may call the above multiplet twisted holomorphic
since 
\eqn\thol{
\left\{
\eqalign{
(\bs_+\oplus\bbs_-)\s=0,\cr
(\bbs_+\oplus\bs_-)\bar\s=0,\cr
}\right.\qquad 
\left\{
\eqalign{
(\bs_+\oplus\bbs_+)\phi_{++} =0,\cr
(\bs_-\oplus\bbs_-)\phi_{--}=0.\cr
}\right.
}

The explicit form of the superalgebra is worked out in the Appendix A. 
The $U(1)_\CR$ charges $(J_L,J_R)_\CR$
and the degrees $(p,q)$ of the various fields are given by
the following tables

\lin{Table 1}

$$

\matrix{
Bose&\phi_{++}&\phi_{--} &\s&\bar\s& B^{2,0} & B^{0,2}&H_0&H^{0,2}&H^{2,0}
          &   A& H^{1,0}&H^{0,1}
\cr
J_L    &0&0&+1& -1 &-1& +1&0&0&0&0&-1&+1
\cr
J_R   &0&0&+1&-1   &+1&-1&0&0&0&0&+1&-1
\cr
p	&1&-1&0&0& -1&+1&0&0&0&0&-1&+1
\cr
q	&1&-1&0&0&+1&-1&0&0&0&0&+1&-1
}
$$

\lin{Table 2}

$$

\matrix{
Fermi&\eta_+& \eta_-&\bar\eta_+ &\bar\eta_-
           &\chi^{2,0}_+&\chi^{2,0}_-&\bar\chi^{0,2}_+&\bar\chi^{0,2}_-
           &\bar\p_+^{1,0}&\bar\p_-^{1,0}&\p_+^{0,1}&\p^{0,1}_-
\cr
J_L    &0 & +1 &0&-1
           &0 & - 1&0&+1
           &- 1& 0 &+1&0
\cr
J_R    &-1&0&+1&0                                    
           &+1&0&-1&0  
           &0&+1&0&-1   
\cr
p	&+ 1&0 &0&-1
	&0&-1&+1&0
	&0&-1&+1&0
\cr
q  	&0&-1&+1&0
	&+1&0&0&-1
	&-1&0&0&- 1     
}
$$

The action functional should have degree $(0,0)$ and $U(1)_\CR$ charge
$(0,0)$ and be annihilated by  all four supercharges.
Those requirements (almost) fix the
action functional in the following form 
\eqn\eaa{
S_4 = \bs_+\bbs_+\bs_-\bbs_-\CF
+\bs_+\bs_-\CW
+\bbs_+\bbs_-\overline\CW 
}
with
\eqn\eab{
\eqalign{
\CF= &\Fr{1}{e^2}\int_X \tr\biggl(
\k F\wedge F
+B^{2,0}\wedge B^{0,2}
-2\s *\bar\s
\biggr),\cr
\CW=&\Fr{1}{e^2}\int_X\tr \left(B^{2,0}\wedge F^{0,2}\right),\cr
}
}
where $\k$ is a K\"{a}hler potential of $X$ and $e$ is the Yang-Mills
coupling constant.
Note that $\bbs_\pm \CW=0$ and $\bs_\pm\overline\CW=0$
which ensures the manifest $N=4$ supersymmetry of the action functional
$S_4$.

Before moving on, we make two comments

\leftskip=\parindent

\noindent
\phantom{T}
{\bf (1)}. Let $\CA$ be the space of all connections on a K\"{a}hler
surface $X$ with K\"{a}hler form $\o=-i\rd\bar\rd \k$.
Using the complex and the K\"{a}hler structures on $X$
we can introduce the corresponding structures on $\CA$.
In particular on $\CA$ we have a K\"{a}her potential
$\tilde \k$
defined via
\eqn\kap{
\tilde\k = \Fr{1}{8\pi^2}\int_{X}\k\tr F\wedge F
}
We assert that our supersymmetry generators can be viewed
as the differentials of Dolbeault equivariant cohomology on
$\CA\oplus\CB$. For instance we have the following relation
\eqn\asuba{
\bs_-^2=0,\qquad
\{\bs_-,\bbs_-\}=\d_{\phi_{--}},\qquad
\bbs_-^2=0.
}
This implies that we have the equivariant K\"{a}hler identity
\eqn\equki{
-i\bs_-\bbs_-\tilde\k
=\Fr{1}{4\pi^2}\int_X\tr\phi_{--} F\wedge\o 
-\Fr{1}{4\pi^2}\int_X\tr\p_-^{0,1}\wedge\bar\p_-^{1,0}\wedge\o,
}
where we used  integration by parts and the Bianchi identity
$\Dp F^{0,2} + \Dpp F^{1,1}=0$. The term proportional to
$\phi_{--}$ is the $\CG$-momentum-map on $\CA$ and the second term
is the K\"{a}hler form on $\CA$. 

\noindent
\phantom{T}
{\bf (2)}. Now let us compare the field contents of our model
with the field contents of  $N=4$ super-Yang-Mills theory.
The latter may be viewed as   
a $N=2$ theory with an adjoint hypermultiplet or as
a $N=1$ theory with three adjoint $N=1$ chiral supermultiplets.
To determine the twisted $N=2$ vetor and hyper multiplets
we have to choose a $N=2$ subalgebra ({\it the Dolbault $\CG$-equivariant
cohomology}). The commutations relations \sac\ implies that
we have four sets of {\it equivalent} $N=2$ subalgebra generated
eigther by  $\bs_+\oplus \bbs_+$, or  $\bs_-\oplus\bbs_-$, or
$\bs_+\oplus\bbs_-$, or $\bs_-\oplus \bbs_+$. For any choice
we can identifiy $\bs$ and $\bbs$, omitting $\pm$ indicies,
with holomorphic and
anti-holomophic differentials, respectively, of Dolbeault
$\CG$-equivariant cohomology;
\eqn\dolequi{
\bs^2=0,\qquad \{\bs,\bbs\}=\d_{\varphi},\qquad\bbs^2=0.
}
In a later section we will perturb the $N=4$ theory
to $N=2$ theory by introducing a bare mass for the hypermultiplet.
In the pure $N=2$ limit any one of the above four cases
lead to the Donalson-Witten theory. 

\noindent
\phantom{T}
In this paper we choose the $N=2$ subalgebra generated
by $\bs_+\oplus \bbs_+$, satisfying the commuation relation \fdfd.
Then
the twisted $N=2$ vector multiplet is given by
\eqn\vectorm{
(A^{1,0},A^{0,1}), (\p^{0,1}_+,\bar\p^{1,0}_+), (\phi, \bar\phi), 
(\eta_-,\bar\eta_-), 
(\chi^{2,0}_-,\bar\chi^{0,2}_-),(H, H^{2,0},H^{0,2}),
}
where $\phi:=\phi_{++}$ and $\bar\phi:=\phi_{--}$.
The remainder corresponds to
the twisted  $N=2$ hypermultiplet:
\eqn\hyperm{
(B^{2,0},B^{0,2}) , (\s,\bar\s),  (\p^{0,1}_-,\bar\p^{1,0}_-), (\eta_+,\bar\eta_+), 
(\chi^{2,0}_+,\bar\chi^{0,2}_+), (H^{1,0},H^{0,1})
}

\leftskip=0pt

\subsec{The Localization}

Now we return to the action functional.
Collecting the terms in the action which depend on the 
auxiliary fields:
\eqn\ead{
\eqalign{
S_4 =&
\Fr{1}{e^2} \int_X\!\! \tr\biggl(
-H^{2,0}\wedge H^{0,2}
-H^{2,0}\wedge F^{0,2}
-H^{0,2}\wedge F^{2,0}
+ H\left(2F\wedge\o -i[B^{2,0},B^{0,2}]\right)
\cr
&
- 2H*H 
+B^{2,0}\wedge\Dpp H^{0,1}
+B^{0,2}\wedge\Dp H^{1,0}
-2H^{1,0}\wedge * H^{0,1}
\biggr)+\ldots
}
}
we can integrate $H^{2,0}, H^{0,2},H^0$
using the following replacement
\eqn\upg{
\eqalign{
H^{2,0}&=F^{2,0},\cr
H\o^2 &= F\wedge\o -\Fr{i}{2}[B^{2,0},B^{0,2}],\cr
H^{0,2}&=F^{0,2},\cr
}
}
Similarly $H^{1,0}, H^{0,1}$ are integrated
out by the replacements 
\eqn\upga{\eqalign{
H^{1,0}= \Fr{1}{2}\Dp^* B^{2,0},\cr
H^{0,1}=\Fr{1}{2}\Dpp^* B^{0,2}.\cr
}
}
The resulting action functional is invariant
under $\bs_\pm$ and $\bbs_\pm$ if we modify
the transformation laws  $(A.3)$, $(A.4)$ and $(A.5)$
in the Appendix
according to \upg\ and \upga.

According  to the fixed point theorem of Witten
the path integral is localized on the fixed point
locus of all four supersymmetries. 
Collecting all the simultaneous fixed points equations
from the transformation laws for fermionic fields
we have
\eqn\uuu{
\eqalign{
&F^{0,2}=0,\cr
&F\wedge\o -\Fr{i}{2}[B^{2,0},B^{0,2}]=0,\cr
&[\bar\s, B^{0,2}]=[\s, B^{0,2}]=[\s,\bar\s]=0,\cr
&\Dpp^*B^{0,2}=d_A\bar\s=0,\cr
}
}
and
\eqn\uuus{
[\phi_{++},\phi_{--}]=0,\qquad
[\phi_{\pm\pm},\bar\s]=0,\qquad
[\phi_{\pm\pm},B^{0,2}]=0,\qquad
d_A\phi_{\pm\pm}=0.
}
These are the Vafa-Witten equations on a K\"ahler manifold
and were derived in a different way in section 2.4 of \VW.
Already at this point we could  exploit 
the  $S^1$-action and the resulting localization pointed out
in \VW. However, we will find that it is more useful to
return to this point in a more general framework.

\newsec{Perturbations to $N=2$ and $N=1$ Theories}

The path integral of any cohomological theory is
localized on the fixed point locus of the global
supersymmetry. We examine how the fixed points
are changed when we perturb the original $N=4$
theory to $N=2$ and $N=1$ theories by introducing
bare masses for the  matter multiplets.
The bare mass for the $N=2$ hypermultiplet will be
introduced by exploiting a   global $S^1$-action  
on  the adjoint hypermultiplet. The path
integral is then localized  on  the fixed point locus
of the unbroken $N=2$ symmetry, which has two branches:
 branch (i) is  the moduli
space of anti-self-dual connections, branch (ii) is  the moduli space
of a special type of abelian Seiberg-Witten monopoles. 
We will further break the  supersymmetry down to $N=1$
to get the perturbed Seiberg-Witten equation presented in
branch (ii). The last step is
crucial in order to achieve  the factorization of the 
Seiberg-Witten basic class.  
The procedure in this section should
be understood in conjunction 
with the more physical approach
taken by Vafa and Witten in section 5 of \VW.

\subsec{Perturbation to $N=2$ Theory}

We want to perturb the $N=4$ theory to a $N=2$
theory maintaining the two supersymmetry generated
by $\bs_+$ and $\bbs_+$.
To begin with we regard the $N=4$ theory as
$N=2$ theory with an adjoint hypermultiplet.
In this language the action functional is written as (compare with \actionnt)
\eqn\asnt{
\eqalign{
&S_4 =\Fr{\bs_+}{e^2}\!\!\int_X\!\tr\biggl(-\chi^{2,0}_-\!\wedge\! F^{0,2}
- B^{2,0}\!\wedge\!\Dpp\p^{0,1}_-\biggr)
+\Fr{\bbs_+}{e^2}\!\int_X\!\tr\!\biggl(-\bar\chi^{0,2}_-\!\wedge\! F^{2,0}
-B^{0,2}\!\wedge\!\Dp\bar\p^{1,0}_-\biggr)
\cr
&
\;+\Fr{\bs_+\bbs_+}{e^2}\!\!\int_X\!\!\tr\!\!\biggl(
\bar\phi\left(2i F\!\wedge\!\o +[B^{2,0},B^{0,2}] -2[\s,\bar\s]\right)
+\chi^{2,0}_-\!\wedge\!\bar\chi^{0,2}_- 
-2\eta_-*\bar\eta_-
-2 \p_-^{0,1}\!\wedge\! *\bar\p^{1,0}_-
\biggr)
}
}

Now we want to perturb the theory to $N=2$ theory
by introducing a bare  mass term to the hypermultiplet.
If we divide the hypermultiplets into 
$$
\eqalign{
\hbox{\bf H}:=(\p^{0,1}_-, H^{0,1},\bar\s,\eta_+,B^{0,2},\bar\chi^{0,2}_+),\cr
\overline{\hbox{\bf H}}:=(\bar\p^{1,0}_-, H^{1,0},\s,\bar\eta_+,B^{2,0},\chi^{2,0}_+),
}
$$
the algebra as well as the action functional are 
invariant under 
the following  infinitesimal $S^1$-action
\eqn\hhh{
\d_m(\hbox{\bf H},\overline{\hbox{\bf H}}) 
= (im \hbox{\bf H}, -i m \overline{\hbox{\bf H}}),
}
where  $m$ is a positive  real parameter.

This $S^1$-action is the unbroken part of the
global  $SO(4)$  symmetry
rotating  the components of the 
two complex bosons  of the untwisted $N=2$ 
hypermultiplet.  It is the right $U(1)_\CR$ symmetry.
Note, from the tables $1$ and $2$ in Sect. $3.2.$, 
that $\hbox{\bf H}$ and $\overline{{\bf H}}$ have $J_R=-1$
and $J_R=+1$ respectively.
We want to extend  the  $\CG$-equivariant
Dolbeault  cohomology to $\CG\times S^1$-equivariant 
one. Then it is natural to assign the degree $(1,1)$
to $m$.
Differentials of the extended equivariant Dobeault cohomology
have  the following commutation relations
\eqn\commmm{
\bs_+^2=0,\qquad
\{\bs_+,\bbs_+\} = \d_{\phi}+ \d_{m},
\qquad
\bbs_+^2=0
}
where $\d_{m}$ is defined in \hhh.

Altogether we thus have 
the following  modified
transformation laws for hypermultiplets,  
denoting $\d = \bar\e_-\bs_+ +\e_-\bbs_+$
\eqn\mntha{
\eqalign{
\d\p^{0,1}_- &=
	\bar\ep_- H^{0,1}
	+\ep_-\Dpp\bar\s,
\cr
\d\bar\p^{1,0}_- &=
	\ep_- H^{1,0}
	+\bar\ep_-\Dp\s,
}\qquad
\eqalign{
\d H^{0,1} =&
	 -\ep_-([\phi,\p^{0,1}_-] +im\p_-^{0,1})
	+\ep_-[\bar\s,\p^{0,1}_+] 
	-\ep_-\Dpp\eta_+,
\cr
\d H^{1,0} =&
	 -\bar\ep_-([\phi,\bar\p^{1,0}_-] -im \bar\p_-^{1,0}) 
	+\bar\ep_-[\s,\bar\p^{1,0}_+] 
	-\bar\ep_-\Dp\bar\eta_+,
}
}
and
\eqn\mnthb{
\eqalign{
\d B^{0,2} 
       =&\ep_-\bar\chi_{+}^{0,2},
\cr
\d B^{2,0}=& \bar\ep_-\chi_+^{2,0} ,
}\qquad
\eqalign{
\d\bar\chi_{+}^{0,2}
     	=& 
	-\bar\ep_-([\phi,B^{0,2}]+im B^{0,2}),
\cr
\d \chi_+^{2,0} =& 
		-\ep_- ([\phi, B^{2,0}]-im B^{0,2}), 
\cr
}
}
and
\eqn\mnthc{
\eqalign{
\d\bar\s &= -\bar\ep_-\eta_+ ,
\cr
\d \s &=  -\ep_-\bar\eta_+,
}
\qquad
\eqalign{
\d\eta_{+}
     &=	+\ep_-([\phi,\bar\s]+im\bar\s)
,\cr
\d\bar\eta_{+}
     &= 	+\bar\ep_-([\phi,\s]-im\s).
\cr
}
}

The action functional is still defined by the formula \asnt,
but using the modified transformations laws \mntha, \mnthb\ and \mnthc,
one obtains  the new expression 
\eqn\ttr{
S_4^\pr = S_4 +\Fr{1}{e^2}\int_X
\tr\left(-i m\phi_{--}\left[B^{2,0},B^{0,2}\right] +2 im\phi_{--}*[\s,\bar \s]
+m\p_-^{0,1}\wedge\bar\p^{1,0}_-\wedge\o
\right),
}
which  contains new $m$-dependent
terms. Physically, these  terms are  part of the 
mass term for the hypermultiplet. To get the remaining
terms we digress briefly.

Recall that the bosonic components of the hypermultiplet
are $\s,\bar\s$ and $B^{2,0},B^{0,2}$.
Now consider the space $\CH$ of all bosonic hypermultiplets.
On $\CH$ we have a $\CG\times S^1$ action. 
Thus we can  introduce a $\CG\times S^1$-invariant  inner product.
We also introduce a compatible complex structure on $\CH$
by declaring $\bar\s$ and $B^{0,2}$ as the holomorphic
coordinates. Now we define  a K\"{a}hler potential on $\CH$
as 
\eqn\kpo{
h = \int_X\tr(B^{2,0}\wedge B^{0,2} -2\s * \bar\s).
}
The relation \commmm\ implies that $\bs_+$ and $\bbs_+$
are the holomorphic and anti-holomorphic 
differentials of  $\CG\times S^1$-equivariant Dolbeault 
cohomology, i.e.,
$$
\eqalign{
\bs_+\s =0,\cr
\bbs_+\bar\s=0,\cr
}\qquad
\eqalign{
\bs_+ B^{0,2}=0,\cr
\bbs_+ B^{2,0}=0.
}
$$
Thus we compute the    equivariant K\"{a}hler form
$I_h$ from the K\"ahler potential $h$ according to
\eqn\exact{
\eqalign{
I_h
&=i\bs_+\bbs_+ h
\cr
&=\int_X\tr\biggl(
i\phi_{++}[B^{2,0},B^{0,2}] 
-2i\phi_{++}[\s,\bar \s] 
+i\chi^{2,0}_+\wedge\bar\chi^{0,2}_+
-2i\bar\eta_+*\eta_+
\cr
&\qquad\qquad
+m B^{2,0}\wedge B^{0,2} -2m\s*\bar \s 
\biggr).
}
}
Note that the terms proportional to $m$ and to
$\phi$ are the momentum maps of the $S^1$ and
$\CG$ actions respectively.
The remaining terms  are  the K\"{a}hler form on 
$\CH$.

The equivariant K\"{a}hler form $I_h$
can now  be used to construct the  promised mass terms, which
we define to be  $\Fr{4m}{e^2}I_\kappa$. Now all elements
of the hypermultiplet have acquired a bare mass.
Putting all the  mass terms together  we  thus 
get the $N=2$ symmetric action
\eqn\rrr{
S_2(m) = S_4^\pr + \Fr{m}{e^2} I_{h},
}
which includes all the bare mass terms for the $N=2$
hypermultiplet \hyperm.

The new $N=2$ supersymmetric
action $S_2(m)$ contains no additional terms in 
$H^{2,0},H^{0,2},H^0$
and $H^{1,0}, H^{0,1}$ compared with $S_4$.
Thus integrating out these fields leads to  the same replacements
as given in  \upg\ and \upga.
With these replacements we can now collect the  fixed point equations
for the unbroken supersymmetry charges
 $\bs_+$ and $\bbs_+$. From the fixed point equations
$\bs_+\bar\eta_-=\bbs_+\eta_-=0$ in (A.3) and
$\bs_+\chi^{2,0}_-=\bbs_+\bar\chi^{0,2}_-=0$ in (A.4)
we have
\eqn\ijhu{
\eqalign{
&F^{0,2} = [\bar\s ,B^{0,2}]=0,\cr
&iF\wedge \o +\Fr{1}{2}[B^{2,0}, B^{0,2}]
            -\Fr{1}{2}[\s,\bar \s]\o\wedge\o=0,\cr
&F^{2,0} =[\s,B^{2,0}] =0.\cr
}
}
From \mntha\
we get
\eqn\ijhuuy{
\eqalign{
&\Dpp^* B^{0,2} =\Dpp\bar\s =0,\cr
&\Dp^* B^{2,0} =\Dp \s =0.\cr
}
}
From  $\bs_+\bar\eta_-=\bbs_+\eta_-=0$ in $(A.3)$
and from
$\bs_+\bar\p^{1,0}_+=\bbs_+\p^{0,1}_+=0$ in $(A.5)$
we have
\eqn\aty{
[\phi,\bar\phi]=0,\qquad d_{\!A}\phi=0.
}
From  \mnthb\ and \mnthc\ we have
\eqn\cson{
\eqalign{
[\phi,B^{0,2}]+im B^{0,2} =0,\cr
[\phi,B^{2,0}]-im B^{2,0} =0,\cr
}\qquad\eqalign{
[\phi,\bar\s]+im \bar\s=0,\cr
[\phi,\s]-im \s=0.\cr
}
}

In studying solutions of  these fixed point equations  we specialize 
to the gauge group $SU(2)$. 
First of all, \aty\
implies that $\phi$ should be diagonalized in the fixed
points. Thus  we have two branches

\noindent
{\bf Branch (i)}:$\phi=0$, i.e.   the gauge symmetry is unbroken.
 Then \cson\ implies that $B^{2,0}, B^{0,2}$ and $\s,\bar \s$
 vanish. So the fixed point equation \ijhu\ reduces to
the  anti-self-duality equation  for the connection $A$: $F_A^+ =0$.

\noindent
{\bf Branch (ii)}: $\phi \neq 0$, i.e. 
 the gauge symmetry is broken down to
$U(1)$. Thus  the bundle $E$ splits into line bundles , 
$E=L\oplus L^{-1}$ with $L\cdot L = -k$,   
and  $\phi=i\phi^a T^a$ takes the form
$\phi =ia T^3$. Then
the only non-trivial solutions of \cson\ 
are, with $m+a= 0$:
\eqn\upb{\eqalign{
B^{0,2}=\left(\matrix{ 0 & \b\cr 0 & 0}\right),\cr
B^{2,0}=\left(\matrix{ 0 & 0\cr \bar\b & 0}\right),\cr
}\qquad
\eqalign{
\bar\s =\left(\matrix{ 0 & \a\cr 0 & 0}\right),\cr
\s=\left(\matrix{ 0 & 0\cr \bar\a & 0}\right).
}
}
Then  \ijhu\ reduces to
\eqn\upc{\eqalign{
F^{0,2}_{L^2}   &=\a\b=0,\cr
iF_{L^2}\wedge\o &= \b\wedge\bar\b -\a\bar\a\o^2,\cr
}\qquad
\eqalign{
\bar\rd_{L^2}^*\b=\bar\rd_{L^2}\a=0.
}
}
Here $\a$ is a section of $L^{2}$ and $\beta$
is a section of $K^{-1}\otimes L^{2}$, with  $K$ denoting
the canonical line bundle. To make progress it 
it is useful to regard the above equation 
as a perturbation of another equation. To achieve this note that
\eqn\cbc{
F_{K^{-1/2}\otimes L^{2}} = \Fr{1}{2}F_{K^{-1}} +F_{L^2} 
\quad
\rightarrow F_{L^2} = F_{K^{-1/2}\otimes L^{2}} 
-\Fr{1}{2}F_{K^{-1}},
}
so that  we can write
\eqn\cbd{
\eqalign{
&F^{0,2}_{L^2} =\a\b=0,\cr
&\Fr{i}{2}F_{\zeta}\wedge \o
= \beta\wedge \bar\beta -\a\bar\a\o^2 
+\Fr{i}{2}F_{K^{-1}}\wedge\o,\cr
}
\qquad
\eqalign{
\bar\rd_{L^2}^*\b=\bar\rd_{L^2}\a=0.
}
}
This is a perturbation of  the Seiberg-Witten
equation \WittenC\ for a $spin^c$ structure 
$\zeta = K^{-1}\otimes L^{4}$; this fact will be crucial
in the next section.
For later use  we also note that  $c_1(\zeta) = w_2(X)$ modulo $2$
since
$c_1(K) = w_2(X)\hbox{ mod } 2$.

\subsec{Perturbation to $N=1$ Theory}

We can further break the remaining $N=2$ symmetry
down to $N=1$ by introducing a  bare mass for
the $N=1$ matter-multiplet. We will do this  preserving
$\bbs_+$-symmetry. Note that, among the $N=2$ vector multiplet
given by \vectorm, the $N=1$ matter multiplet consists
of $(\p^{0,1}_+,\phi,\bar\phi,\bar\eta_-,\chi^{2,0}_-)$.

The  required  mass term involves an  holomorphic two-form
$ \o^{2,0}$ and has the form
\eqn\uua{
\Fr{1}{e^2}I_{\o^{2,0}}=
\Fr{1}{2 e^2}\int_X\tr(\p_+^{0,1}\wedge\p_+^{0,1})\wedge\o^{2,0}.
}
This term  is invariant under $\bs_+$-symmetry,  but  not   invariant
under the $\bbs_+$ symmetry;
\eqn\vfg{
 \Fr{\bbs_+}{e^2}I_{\o^{2,0}}=-\Fr{1}{e^2}\int_X\tr \phi
\Dpp\p^{0,1}_+\wedge\o^{2,0}.
}
However, the imposition of the
$\chi^{2,0}_-$ equation of motion leads to invariance.
The relevant term in the action $S_2(m)$ is
$-\int_X\tr \chi^{2,0}_-\wedge\Dpp\p^{0,1}_+$.
If we  add \uua\ to the action $S_2(m)$ of \rrr\ and 
at the same time change the  $\bbs_+$-transformation 
of $\chi^{2,0}_-$ according to
\eqn\kji{
\bbs_+\chi^{2,0}_- =[\bar\s,B^{2,0}] \longrightarrow
\bbs_+\chi^{2,0}_- = [\bar\s,B^{2,0}]-\phi\o^{2,0},
}
then  the new action $S_2(m)^\pr + \Fr{1}{e^2}I_{\o^{2,0}}$
enjoys  $\bbs_+$-symmetry. Here $S_2(m)^\pr$ is given by
\eqn\ntpr{
S_2(m)^\pr = S_2(m) -\Fr{1}{e^2}\int_X\tr \phi[\s,B^{0,2}]\wedge\o^{2,0},
}
where the additional term is due to the modification \kji\ (see \asnt).
Since $\bbs_+\phi=0$, we still have the property $\bbs_+^2=0$.
In this way  the one component $\p_+^{0,1}$ of the $N=1$
chiral superfield has obtained  a mass.
To  give mass to  the remaining components
in the $N=1$ matter multiplet we  add 
the following $\bbs_+$-exact terms to the action
\eqn\rewa{
\eqalign{
I_{\phi\bar\phi}= &
-\bbs_+\int_X\tr\left(\bar\phi\chi^{2,0}_-\right)
\wedge\o^{0,2}
\cr
= &
-\int_X\tr\left(\bar\phi[\bar\s, B^{2,0}]\right)\wedge\o^{0,2} 
+\int_X\tr\left(\phi\bar\phi\right)\o^{2,0}\wedge\o^{0,2}
\cr
&
-\int_X\tr\left(\bar\eta_-\chi^{2,0}_-\right)
\wedge\o^{0,2}
}
}
A similar  prescription for breaking  pure $N=2$ theory
down to $N=1$  was given by Witten in \WittenB.

To sum up, the total action 
\eqn\tracx{
S_1(m,\o^{0,2}) = S_2(m)^\pr 
+ \Fr{1}{e^2}I_{\o^{0,2}}+\Fr{1}{e^2}I_{\phi\bar\phi},
}
has only $\bbs_+$ supersymmetry and
all the matter multiplets have a  bare mass.

Now   the fixed point equations  \ijhu\ 
receive an important change due to the
modification of the $\bbs_+$ transformation law
of $\chi^{0,2}_-$ given by \kji.
The new fixed point equations are
\eqn\rma{\eqalign{
F^{0,2} =[\bar\s ,B^{0,2}]-\phi\o^{0,2}=0,\cr
iF\wedge \o +\Fr{1}{2}[B^{2,0}, B^{0,2}]
            -\Fr{1}{2}[\s,\bar \s]\o\wedge\o &=0,\cr
}\qquad
\eqalign{\Dpp^* B^{0,2}=\Dp\bar\s=0,}
}
while \aty\ and \cson\ remain unchanged. Thus there are again
two branches and 
it is easy to see that the fixed point equation
for branch (i) is unchanged, while 
the equations for branch (ii) become:
\eqn\upcd{\eqalign{
F^{0,2}_\zeta   &=
\a\b-m\o^{0,2}=0,\cr
\Fr{i}{2}F_\zeta\wedge\o &= \b\wedge\bar\b -\a\bar\a\o\wedge\o
-\Fr{i}{2}F_{K^{-1}},\cr
}
}
where $\zeta = K^{-1}\otimes L^4$.
This  is  a perturbed version of the Seiberg-Witten equation,
containing the perturbation
introduced by Witten in \WittenC. 
The condition
\eqn\fact{
\a\b=m\o^{0,2},
}
gives  the crucial factorization condition
of  the Seiberg-Witten basic classes.

\newsec{Analysis of Branch (ii)}

\subsec{A Selection Rule}

Our analysis of branch (ii) exploits the relation of the 
 the defining  equations  with the 
the Seiberg-Witten equation. 

As a first step we 
need to classify which Seiberg-Witten classes contribute
to branch (ii). 
For an arbitrary $spin^c$ structure $x$,
which can  always be written in terms of an arbitrary integral line bundle
 $\xi$  as
\eqn\ecbm{
x=K^{-1}\otimes \xi^{2},
}
we have  an associated Seiberg-Witten equation.
If the square root $\xi^{1/2}=L$ of $\xi$ exists, the
Seiberg-Witten equation is identical to 
the fixed point equation \cbd\ of branch (ii). 
The inclusion of the  perturbation in \upcd\ further implies that
we also have to satisfy the  factorization condition 
$\a\b =\o^{0,2}$, where we have 
scaled  $m=1$ in \fact. 

Let the canonical divisor $K$ be 
given by $K=\sum_i r_i C_i$, where the $C_i$  are  irreducible
components. The factorization means that
\eqn\facto{
K^{1/2}\otimes x^{1/2}= \xi 
=\sum_i s_i C_i,
}
where $s_i$ are integers with $0\leq s_i\leq r_i$
and $I$ is the trivial line bundle.
Thus, the question  of which Seiberg-Witten classes contribute
to branch (ii) reduces to finding 
 line bundles  $L$ satisfying  $L\cdot L =-k$ and 
\eqn\ebd{
2L =\sum_{i=1}^n s_i C_i, \qquad 0\leq s_i\leq r_i.
}
Now let $x$ be a Seiberg-Witten basic class. If $(x^{1/4}\otimes K^{1/4})$
exists as a line bundle, then the associated SW invariant
$n_x$ contributes to the path integral in branch (ii).
Note that $(x^{1/2}\otimes K^{1/2})$ always exist
as a line bundle. The question is thus whether the square root of
 $(x^{1/2}\otimes K^{1/2})$ exists, which is the case iff
\eqn\ebe{
\Fr{1}{2}[x + K] =  0,
}
or, 
equivalently
\eqn\ebee{
\Fr{1}{2}[x + w_2(X)] = 0,
}
where $[...]$ means the mod $2$ reduction.
Here $w_2(X)$ is the second Stiefel-Whitney class
of our K\"{a}hler manifold $X$.
In the $SU(2)$ case such a  square root may not exist. 
However, if we repeat the analysis for 
an $SO(3)$  bundle $E$, 
the factorization condition  can be met provided 
the  second Stiefel-Whitney class
$w_2(E)$  of $E$ satisfies
\eqn\ebf{
\Fr{1}{2}[x + w_2(X)] \equiv w_2(E).
}

With the abbreviations $z_0 = w_2(X)$,
 $z= w_2(E)$ and
$2 x^\pr= x + K$ the branch (ii) contribution has the form
\eqn\delll{
\sum_{x}n_x \d_{z,[x^\pr]}\times(\ldots),
}
where the summation is over all Seiberg-Witten
basic classes $x$. This general form  applies to both  the $SU(2)$ and 
the $ SO(3)$ case. In principle one could proceed to compute the 
 branch (ii) contribution directly using localization techniques.
However, in practice this requires that one starts with a suitable 
compactification of the moduli space of the Vafa-Witten equations in order
to make the integration over the normal bundle of  branch (ii)
 well-defined. Here we are not able to follow that path. Instead we 
determine the branch (ii) contribution to the partition function by
a generalisation of the results of  Vafa and Witten in \VW.

\subsec{Partition Function for Branch (ii)}

Consider the case  $K =\sum_{i}^n [C_i]$
where the  $[C_i]$  are irreducible and disjoint.
Vafa and Witten made a  prediction for
what we call the contribution to the partition function
from branch (ii) (eq. 5.50 in \VW). For the $SU(2)$ case
the answer is
\eqn\cci{
\left(\Fr{G(q^2)}{4}\right)^{\n/2}
\left(\Fr{\th_0}{\eta^2}\right)^{\sum_{i=1}^n (1-g_i)}
\sum_{\vec{\e}} \d_{0,w_2(\vec{\e})}
\left(\prod_{i=1}^n t_i^{\e_i}\right)
\left(\Fr{\th_1}{\th_0}\right)^{\sum \e_i(1-g_i)},
}
where $\vec{\e}=(\e_1,\ldots,\e_n)$ and $\e_i=0$ or
$1$ chosen independently. 
Here $\n =(\chi+\s)/4$
and $\chi$ and $\s$ denote the Euler number and
signature of the manifold, respectively.

{}From our perspective the sum and the  delta function 
$\d_{0,w_2(\vec{\e})}$ can be understood as follows.
 What is called 
$w_2(\vec{\e})$ in \VW\ is a special
form of $[x^\pr]$, so that the sum in \cci\
is over the same range as the sum \delll.
In our notation, the factorization condition has
the form
\eqn\uijo{
x^\pr = 2L =\sum_{i=1}^{n}\e_i[C_i],\qquad 0\leq s_i\leq 1.
}
{}From
\eqn\cce{
k = -L\cdot L =-\Fr{1}{4}x^\pr\cdot x^\pr
= -\Fr{1}{4}\sum_i s^2_i(g_i -1)
}
and since $s^2_i = s_i$ for $s_i=0$ or $1$,
we recover the formula 
\eqn\ccf{
k = -\Fr{1}{4}\sum_i \e_i(g_i -1),\qquad
0\leq \e_i\leq 1.
}
given in \VW.
Note also that $\sum_{i=1}^n(g_i-1)=K\cdot K=2\chi + 3\s$.

We now  propose a  formula for  the branch (ii)
contributions on general  K\"ahler manifolds with $b_2^+ \geq 3$.
We replace
\eqn\taa{
\sum_{\vec{\e}}\d_{0,w_2(\vec{\e})}
\left(\prod_{i=1}^n t_i^{\e_i}\right)
\left(\Fr{\th_1}{\th_0}\right)^{\sum \e_i(1-g_i)}
\longrightarrow
(-1)^\n\sum_{x}\d_{0, [x^\pr]}
n_x
\left(\Fr{\th_1}{\th_0}\right)^{-x^\pr\!\cdot x^\pr},
}
where the summation is over all Seiberg-Witten basic
classes $x$ with the Seiberg-Witten invariants $n_x$.
Then, \cci\ can be written as
\eqn\tac{
(-1)^\n\left(\Fr{G(q^2)}{4}\right)^{\n/2}
\left(\Fr{\th_0}{\eta^2}\right)^{-2\chi -3\s}
\sum_{x}\d_{0,x^\pr}n_x
\left(\Fr{\th_1}{\th_0}\right)^{-x^\pr\!\cdot x^\pr}.
}
In the $SO(3)$ case, for a fixed $z=w_2(E)$, we immediately get
\eqn\tad{
(-1)^\n\left(\Fr{G(q^2)}{4}\right)^{\n/2}
\left(\Fr{\th_0}{\eta^2}\right)^{-2\chi -3\s}
\sum_{x}\d_{z,x^\pr}n_x
\left(\Fr{\th_1}{\th_0}\right)^{-x^\pr\!\cdot x^\pr}.
}

\newsec{The Partition Function}

Vafa and Witten make  a precise  statement  about the  expected 
behaviour of  the  partition function of $N=4$ super-Yang-Mills 
theory  under $S$-duality \VW.
According to \VW\  the partition
function of $N=4$ theory is a modular form invariant
under the $\G_0(4)$ subgroup of $SL(2,\msbm{Z})$.
If this is true  the total partition function
can be determined 
from the contribution \tad\ of branch (ii) alone, as we shall now 
show. The  basic idea is  to  examine 
 the terms generated by applying the 
$S$-duality  transformations  corresponding to 
   $\t \rightarrow -1/\t$ 
and  $\t \rightarrow \t +1$ to \tad. Combining the 
resulting  terms in a convenient fashion one gets
\eqn\guessc{\eqalign{
&Z_z = 
(-1)^\n\left(\Fr{G(q^2)}{4}\right)^{\n/2}
\left(\Fr{\th_0}{\eta^2}\right)^{-2\chi -3\s}
\sum_{x}\d_{z,[x^\pr]}n_x
\left(\Fr{\th_1}{\th_0}\right)^{-x^\pr\!\cdot x^\pr}
\cr
&+ 2^{1-b_1} 
\left(\Fr{G(q^{1/2})}{4}\right)^{\n/2}
\left(\Fr{\th_0 +\th_1}{2\eta^2}\right)^{-2\chi -3\s}
\sum_{x}(-1)^{[x^\pr]\cdot z} n_x 
\left(\Fr{\th_0 - \th_1}{\th_0+\th_1}\right)^{ 
-x^\pr\!\cdot x^\pr}
\cr
&
+ 2^{1-b_1} i^{-z^2}
\left(\Fr{G(-q^{1/2})}{4}\right)^{\n/2}
\left(\Fr{\th_0 -i\th_1}{2\eta^2}\right)^{-2\chi -3\s}
\sum_{x} (-1)^{[x^\pr]\cdot z}n_x 
\left(\Fr{\th_0+ i\th_1}{\th_0 -i\th_1}\right)^{ 
-x^\pr\!\cdot x^\pr},
}
}
 where the sum 
$\sum_{x}$
is  over all Seiberg-Witten basic classes, as before.
In principle there could be a  contribution to the partition
function which can not be obtained by performing
modular transformations of the contribution of branch (ii).
However, such a contribution vanishes for
a manifold with $b_2^+ > 1$.

According to \VW\  the required transformation behaviour
under  $S$-duality is:
\eqn\sduality{
Z_y(-1/\tau) = 2^{-b_2/2}(-1)^\n
\left(\Fr{\tau}{i}\right)^{-\chi/2}
\sum_{z}(-1)^{z\cdot y}Z_z(\tau)
}
We can check that our proposed expression \guessc\ transforms
correctly as follows.
First we insert \guessc\ into the RHS of \sduality\ and 
obtain
\eqn\rhs{
\eqalign{
& RHS = \left(\Fr{\tau}{i}\right)^{-\chi/2}\biggl[
2^{-b_2/2}
\left(\Fr{G(q^2)}{4}\right)^{\Fr{\n}{2}}
\left(\Fr{\th_0}{\eta^2}\right)^{-2\chi -3\s}
\sum_{x}(-1)^{z\cdot [x^\pr] }n_x
\left(\Fr{\th_1}{\th_0}\right)^{-x^\pr\!\cdot x^\pr}
\cr
&+ 2^{1-b_1+b_2/2}(-1)^\n
\left(\Fr{G(q^{\Fr{1}{2}})}{4}\right)^{\Fr{\n}{2}}
\left(\Fr{\th_0+\th_1}{2\eta^2}\right)^{-2\chi -3\s}
\sum_{x}\d_{z,[x^\pr]} n_x 
\left(\Fr{\th_0- \th_1}{\th_0+\th_1}\right)^{     
-x^\pr\!\cdot x^\pr}
\cr
&
+ 2^{1-b_1}(-1)^\n  i^{z^2 -\Fr{\s}{2}}
\!\left(\Fr{G(-q^{\Fr{1}{2}})}{4}\right)^{\Fr{\n}{2}}
\!\!\left(\Fr{\th_0 -i\th_1}{2\eta^2}\right)^{\!-2\chi -3\s}
\!\sum_{x} (-1)^{[x^\pr] \cdot z}i^{x^\pr\!\cdot x^\pr}n_x 
\!\left(\Fr{\th_0+ i\th_1}{\th_0-i\th_1}\right)^{
-x^\pr\!\cdot x^\pr}
\biggr]
}
}
where we used
\eqn\ssumr{
\eqalign{
&\sum_{z}(-1)^{z\cdot y} \d_{z,x^\pr}
=(-1)^{y\cdot x^\pr},\cr
&\sum_{z}(-1)^{z\cdot y + z\cdot x^\pr}
= 2^{b_2}\d_{y,x^\pr},\cr
&\sum_{z}(-1)^{z\cdot y} i^{-z^2}(-1)^{z \cdot x^\pr}
= 2^{b_2/2}i^{+y^2 -\s/2 + x^\pr\!\cdot x^\pr}(-1)^{x^\pr\cdot y}.
}
}
Carefully taking into account differences in notation these 
formulae follow from those noted as eq. (5.40) in \VW.
In comparing \rhs\ with the expression \guessc\ evaluated 
at $-1/\t$ one finds that the first line  and second line in \rhs\  
equal, respectively,  the second and first line in \guessc\ evaluated
at $-1/\t$. The third line of \rhs\ should thus be compared with
the third line in \guessc\ at $-1/\t$. The equality
here may  require some  explanation.
Performing $\t \rightarrow -1/\t$ in  the third line on \guessc\
one  finds,  with some
rearrangements, that 
\eqn\yyu{\eqalign{
2^{1-b_1}(-1)^\n i^{z^2-\s/2}\left(\Fr{\t}{i}\right)^{-\chi/2}
&\left(\Fr{G(-q^{1/2})}{4}\right)^{\n/2}
\left(\Fr{\th_0 -i\th_1}{2\eta^2}\right)^{-2\chi -3\s}
\cr
&\times
\sum_{x} (-1)^{-z^2 +[x^\pr]\cdot z} 
i^{-x^\pr\!\cdot x^\pr}i^\s (-1)^\n 
n_x 
\left(\Fr{\th_0- i\th_1}{\th_0 +i\th_1}\right)^{ 
-x^\pr\cdot x^\pr + 2\chi + 3\s}.
}
}
We want to show that the above is identical to 
the third term in \rhs. 
A crucial property is that $-x$ is  a Seiberg-Witten basic
class if $x$ is.
Note  also that $x^\pr = \Fr{1}{2}x + \Fr{1}{2}K$.
 Writing  $\bar x^\pr =-\Fr{1}{2}x + 
\Fr{1}{2}K$ we have 
$-x^\pr\cdot x^\pr + 2\chi +3\s = \bar x^\pr\cdot \bar x^\pr$,
since a Seiberg-Witten basic class
$x$ satisfies $x\cdot x =2\chi+3\s$.
Now the second line of \yyu\ can be rewritten as
\eqn\yyv{
\sum_{-x} (-1)^{-z^2 + x^2}(-1)^{[\bar x^\pr]\cdot z}
i^{\bar x^\pr\!\cdot\bar x^\pr} n_{-x} 
\left(\Fr{\th_0+i\th_1}{\th_0 -i\th_1}\right)^{ 
-\bar x^\pr\cdot \bar x^\pr},
}
where we used $n_{-x} = (-1)^\n n_x$ and the fact that
$\n =(\chi+\s)/4$ is an integer.
The Wu formula implies $(-1)^{-z^2 + x^2} =1$, and we 
replace the dummy variable $-x, \bar x^\pr$ with $x,x^\pr$
to complete the proof.

\subsec{A Relation with Strings}

Taubes proved that Seiberg-Witten
invariants (SW)  are equivalent to Gromov-Witten invariants
(Gr)
for a symplectic $4$-manifold of simple type \Taubes. 
Here we  only consider a K\"{a}hler surface. 
Let $\xi$ be a non-trivial, complex line bundle over
$X$ and use $\xi$ to define a $spin^c$ structure
$x=K^{-1}\otimes \xi^2$. Then
$SW(K^{-1}\otimes \xi^{2})= Gr(\xi)$.
Consider a line bundle $\xi$ such that 
 $SW(K^{-1}\otimes \xi^{2})\neq 0$, then
the Poincar\'{e} dual of  $c_1(\xi)$ is represented by the 
fundamental class of an embedded, holomorphic submanifold 
with, say,  $n$  irreducible components. Then each component
$H_i$ satisfies the adjunction formula 
$g(H_i)= 1 + H_i\cdot H_i$, where $g(H_i)$ is the genus of $H_i$.
We can define  the  integer multiplicities  $a_i$
by writing 
$\xi = \sum_{i=1}^n a_i H_i$.

Let the canonical divisor $K$ be given by
a union of irreducible components $C_i$ with
multiplicities  $r_i$, i.e 
$K =\sum_i r_i C_i$.
The factorization of a Seiberg-Witten basic class $x$ means
that
\eqn\cbt{
K^{1/2}\otimes x^{1/2}
= \xi
= \sum_i s_i C_i
}
where  the $s_i$ are integers with $0\leq s_i\leq r_i$.
Consequently, Taubes' result leads to the  
identifications
\eqn\uuiop{
a_i=s_i,\qquad  C_i= H_i.
}
Physically speaking this means  that the 
world sheets  of the superconducting
cosmic strings discussed by Witten  in \WittenB\
are embedded holomorphic curves.

Recall that for a fixed instanton
number $k$  the Seiberg-Witten classes  $x=-K + 2x^\pr$,
with $x^\pr\cdot x^\pr = -4k$ and $z=[x^\pr]$,
contribute to the partition function of $N=4$
theory in branch (ii). From the above discussion
we identify $x^\pr$ with holomorphic
curve $x^\pr = \sum_{i=1}^n s_i H_i$ and
$1- g(x^\pr) = -x^\pr\cdot x^\pr$. So the branch (ii)
contribution can be written as the sum of contributions
of all holomorphic curves $\S$ with $[\S]= z$. The summation
over the (space-time) instanton numbers is replaced with
the summation over the genus of the holomorphic
curves (the world-sheet instantons).
So our formula \guessc\   for the partition function of $N=4$ theory 
 can  also be viewed as  a genus expansion:
\eqn\guesse{\eqalign{
&Z_z = 
(-1)^\n\left(\Fr{G(q^2)}{4}\right)^{\Fr{\n}{2}}
\left(\Fr{\th_0}{\eta^2}\right)^{-2\chi -3\s}
\sum_{\S}\d_{z,[\S] }Gr(\S)
\left(\Fr{\th_1}{\th_0}\right)^{1-g(\S)}
\cr
&+ 2^{1 -b_1}
\left(\Fr{G\left(q^{1/2}\right)}{4}\right)^{\Fr{\n}{2}}
\left(\Fr{\th_0+\th_1}{2\eta^2}\right)^{-2\chi -3\s}
\sum_{\S}(-1)^{[\S]\cdot z}Gr(\S) 
\left(\Fr{\th_0 - \th_1}{\th_0 +\th_1}\right)^{ 
1-g(\S)}
\cr
&
+ 2^{1 -b_1}
i^{-z^2}\left(\Fr{G\left(-q^{1/2}\right)}{4}\right)^{\Fr{\n}{2}}
\left(\Fr{\th_0-i\th_1}{2\eta^2}\right)^{-2\chi -3\s}
\!\!\sum_{\S}(-1)^{[\S]\cdot z}Gr(\S) 
\left(\Fr{\th_0+ i\th_1}{\th_0 -i\th_1}\right)^{ 
1-g(\S)}.
}
}

\subsec{The $N=2$ Limit and Donaldson-Witten Invariants}

It is instructive to rewrite the 
formula \guessc\
as follows:
\eqn\guessd{\eqalign{
&Z_z = 
(-1)^\n\left(\Fr{G(q^2)}{4}\right)^{\Fr{\n}{2}}
\left(\Fr{\th_0}{\eta^2}\right)^{-2\chi -3\s}
\sum_{x}\d_{z,[x^\pr] }n_x
\left(\Fr{\th_1}{\th_0}\right)^{-x^\pr\!\cdot x^\pr}
\cr
&+ 2^{1 -b_1+ \Fr{1}{4}(7\chi +11\s)} 
G\left(q^{1/2}\right)^{\Fr{\n}{2}}
\left(\Fr{\th_0+\th_1}{\eta^2}\right)^{-2\chi -3\s}
\sum_{x}(-1)^{[x^\pr]\cdot z}n_x 
\left(\Fr{\th_0 - \th_1}{\th_0 +\th_1}\right)^{ 
-x^\pr\!\cdot x^\pr}
\cr
&
+ 2^{1 -b_1 + \Fr{1}{4}(7\chi +11\s)}
i^{-z^2}G\left(-q^{1/2}\right)^{\Fr{\n}{2}}
\left(\Fr{\th_0-i\th_1}{\eta^2}\right)^{-2\chi -3\s}
\sum_{x}(-1)^{[x^\pr]\cdot z}n_x 
\left(\Fr{\th_0+ i\th_1}{\th_0 -i\th_1}\right)^{ 
-x^\pr\!\cdot x^\pr}
}
}
The fact that this formula can naturally be grouped into
three terms whereas we classically  think of contributions
from two branches can be understood physically as follows. 
The first  term  is  the contribution from 
branch (ii) and stems from the singularity in the $u$-plane 
due to the massless adjoint hypermultiplet.  
The remaining two terms are the contribution
from branch (i), which classically corresponds to the 
singularity at the origin of the $u$-plane. 
Geometrically,  the branch (i) contribution
is the Euler characteristic of the moduli space of instantons \VW. 
The fact that this 
contribution is made up from two terms is due to a quantum effect:
 the classical singularity at 
the origin of the $u$-plane bifurcates quantum mechanically \SWb.

{}From the  above formula  we can  recover
the Donaldson invariants for gauge groups $SU(2)$ and
$SO(3)$ as follows. For a simply connected manifold of simple type,
Witten's  formula for the generating functional of 
Donaldson's invariants is \WittenC
\eqn\witten{
\eqalign{
\left<e^{\hat v +\l u}\right>_z
=&
2^{1 + \Fr{1}{4}(7\chi +11\s)}\exp(v^2/2 +2\l) 
\sum_{x}(-1)^{[x^\pr]\cdot z}n_x e^{v\cdot x}
\cr
&
+2^{1 + \Fr{1}{4}(7\chi +11\s)}i^{\n-z^2}\exp(-v^2/2 -2\l) 
\sum_{x}(-1)^{[x^\pr]\cdot z}n_x e^{-iv\cdot x}.
}
}
Here $\hat v$ is the observable $\CO_2$ \ake\
associated with a two-dimensional cycle $v$ and $u=\CO_4$.
The expectation value is computed  using (twisted) $N=2$
super-Yang-Mills theory.

To obtain the above formula from the $N=4$ theory
we could turn on the observables \ake\ after breaking the  supersymmetry
down to $N=2$ and, following \SWb\SWa, take  the double scaling limit
$m\rightarrow \infty$ and $q\rightarrow 0$ with $\L^2=2q^{1/2}m^2$
being fixed. 
In this limit the singularity coming from
the massless adjoint hypermultiplet (branch (ii))
moves to infinity in the $u$-plane and no longer contributes to
the path integral. On the other hand the two other singularities
remain at the points $u= \pm \L^2$ (in
 Donaldson theory $\L^2$ is  normalized to $2$). 
Here we  are not able to consider general expectation values of 
observables.
However, we can compute the  $N=2$  limit  of the partition function
   \guessd\ (the $q\rightarrow 0$ limit since \guessd\ does not 
depend on $m$).
The leading terms  only come from the  second and the third  lines  
and are given by
\eqn\uiui{
2^{1-b_1 + \Fr{1}{4}(7\chi +11\s)}\biggl( 
\sum_{x}(-1)^{[x^\pr]\cdot z}n_x 
+i^{\n-z^2} 
\sum_{x}(-1)^{[x^\pr]\cdot z}n_x \biggr)
\left(q^{3\n/4} +\ldots\right)
}
Note that this  partition function vanishes unless
the dimension of the moduli space of instantons
is zero. Since $dim_\msbm{C}\CM_k = 4k -3\n$, this occurs
when $k = 3\n/4$, thus explaining  the leading term
$q^{3\n/4}$ in \uiui. In fact,  the expression
\uiui\ contains all the non-trivial information
about Donaldson's invariants.

Recently an important contribution to
related issues appeared in  \MW.
It would be fruitful to apply the physical methods of \MW\
to the problems addressed  here. In that way one 
could  compute the entire generating functional
of the $N=2$ theory with a massive adjoint hypermultiplet, and 
work on   more general four-manifolds.

The relation between the $S^1$ action and 
the mass term of the hypermultiplet described in 
 this paper were summarized and  used
by one of us (JSP) in  the  paper \Monads. There
the same sequence of perturbations $N=4,2,1$ was used 
to  relate  the zero-dimensional reduction of the   Vafa-Witten
equation $(N=4)$  to the ADHM description of instantons $(N=2)$
and of torsion-free-sheaves $(N=1)$. Subsequently
the same $S^1$-action and its application to the mass
perturbation of the $N=4$
theory was  also considered in  \LL.

\newsec{Comments on Sigma-Model Approach}

In this paper we formulated a twisted $N=4$ SYM theory on
a K\"{a}hler manifold as a topological gauge theory
based on balanced equivariant Dolbeault cohomology. 
It turns out that the algebra of balanced Dolbeault cohomology
is isomorphic to the algebra of physical $N_{ws}=(2,2)$ supersymmetric
gauge theory in two-dimensions.  The internal symmetry of
our model includes $SO(1,1)$  such that
the indicies $\pm$ in $\bs_\pm$ and $\bbs_\pm$ can be
identified with the $SO(1,1)$ spinor indecies.
Then we may have a natural  extension of the model
to live in a product manifold $X\times \S$, where the $SO(1,1)$ 
acts on the two-dimensional surface $\S$ as the Lorentz symmetry. 
The two equivariant differentials $\bs_+$ and $\bbs_+$ can be
identified with the two left-moving supercharges on $\S$.
The other two differentials $\bs_-$ and $\bbs_-$ are identified with
the two right-moving supercharges on $\S$. With the above extension
$\phi_{\pm\pm}$ correspond to the components of two-dimensional
vector in the light coordinate (or in the complex coordinate).

Now the commutations relations \sac\ are modified as 
\eqn\tsac{
\eqalign{
\{\bs_+,\bbs_+\} = -\rd_{++}+\d_{\phi_{++}},\cr
\{\bs_-,\bbs_-\} = -\rd_{--}+\d_{\phi_{--}},\cr
}\qquad
\eqalign{
\{\bs_+,\bbs_-\}=\d_\s,\cr 
\{\bs_-,\bbs_+\} = \d_{\bar\s},\cr
}
\qquad
\eqalign{
\{\bs_+,\bs_-\}=0,\cr
\{\bbs_+,\bbs_-\}=0.\cr
}\qquad
\eqalign{
\bs_\pm^2=0,\cr
\bbs_\pm^2=0.
}
}
Consequently we may intepret the $N_{ws}=(2,2)$ supercharges 
in two-dimensions as the differentials of the balanced
$P_\S\times \CG$-equivariant Dolbeault cohomology, where
$P_\S$ denotes the group of  translation along $\S$.
The supersymmetry transformation laws are obtained
from those in Appendix A
by replacing $\phi_{\pm\pm}$ with the two-dimensional covariant
derivatives $D_{\pm\pm}$. Now we identify $\bar\ep_-$ and $\ep_-$
with sections of $K^{-1/2}_\S$ and $\bar\ep_+$ and $\ep_+$
with sections of $\bar K^{-1/2}_\S$, where $K_\S$ denotes the
canonical line bundle on $\S$. Those generators are scalars on $X$.
The balanced structure is simply the classical chiral symmetry. 
The $U(1)_\CR$ symmetry in Sect.~$3.1$ and $3.2$, with the left
and right charges in \rcharge\  and Tables $1$ and $2$,
becomes the $\CR$ symmetry of the $N_{ws}=(2,2)$ theory.
The two holomorphic quartets in \ffa\ correspond to two
$N_{ws}=(2,2)$ chiral matter multiplets. The consistent
nine-plet in \ffcccc\ corresponds the $N_{ws}=(2,2)$
vector multiplet.

The action functional is defined as in \eaa\ with a slight modification
as follows
\eqn\teaa{
S = \bs_+\bbs_+\bs_-\bbs_-\int_\S d\m \CF
+\bs_+\bs_-\int_\S d\m \CW
+\bbs_+\bbs_-\int_\S d\m \overline\CW, 
}
where $d\m$ is the two-dimensional volume-form. 
One may identify $\CW$ with chiral superpotential.
Maintaining the $U(1)$ $\CR$ symmetry
we might consider more general action functional 
\eqn\teaa{
S(t,\bar t) = S + t\bbs_+\bs_-\int_{X\times\S} d\tilde \m\tr \s
+\bar t\bs_+\bbs_-\int_{X\times\S} d\tilde\m\tr \bar\s
}
with $t = \Fr{\th}{2\pi} + i r$. The additional term corresponds to
the Fayet-Iluopolos term (if we have a $U(1)$ factor in the gauge
group).

The resulting theory may be viewed 
as an infinite dimensional $\CN_{ws}=(2,2)$ supersymmetric 
gauged linear-sigma model \GLSM\Wittengr.  
Since our model restricted to $X$ is a topological
theory the only scale dependence is from two dimensions.
Picking the volume of $X$ being small and taking the infrared limit
$vol(\S)\rightarrow \infty$ in two-dimensions our model flows
to  a non-linear sigma-model which target space is the
Vafa-Witten moduli space on $X$.
Thus we may practically regard our extended model  a two-dimensional
$N_{ws}=(2,2)$ theory. 

We may regard the sigma-model viewpoints as a unifying 
framework for the four-dimensional theory.
By considering topological sigma models \Wittenmirror,
we have string theoretic generalization of 
Donaldson-Witten and Vafa-Witten theories on a K\"{a}hler surface.
The Donaldson-Witten theory would be viewed 
as an low-energy effective space-time theory 
for the twisted version, i.e., $A$-model, of our sigma model.  
The Vafa-Witten theory we studied in this paper
would be viewed as an effective space-time theory which computes the
Witten index of world-sheet supersymmetry.  We may also
embedd the Vafa-Witten theory into  the half-twisted
version of our model. The partition function of the half-twisted
model is the elliptic genus which is the index of $\bbs_+$.\foot{Note
that the constructive definition of Vafa-Witten theory involves only 
one supersymmetry, say $\bbs_+$. We may identify the partition
function of Vafa-Witten theory as the degeneration of the index 
of two-dimensional supercharge
$\bbs_+$ to the contributions of the constant modes in two-dimensions. 
} 
At present we do not know if further stringy corrections
involve new differential-topological information on a four-manifolds beyond 
the Seiberg-Witten invariants.

The sigma model approach for K\"{a}her
manifolds with $b_2^+(X)=1$ with $b_2(X) < 10$ and $b_2^+(X)=3$ would be 
very interesting. In the both cases we have an important
vanishing theorem that the Vafa-Witten equation reduces to the equations
for Yang-Mills instantons \VW. Then our model in the infrared limit
flows to supersymmetric non-linear sigma-model where target
space is the moduli space of instantons on $X$. 
For  $X$ is a $K3$ surface ($b_2^+(X)=3$) our model
is closely related to the Higgs branch of the $D_1-D_5$ system where
$D_5$ branes wrapp around $X$.  We expect that our model
flows to  $N_{ws}=(4,4)$ superconformal theory in the infrared limit. 
According to a celebrated conjecture of Maldacena the superconformal
theory  is dual to type $IIB$ superstrings on $AdS_3\times S^3\times K3$ \juanLN.

\def\ack{\bigbreak\bigskip\bigskip\centerline{
{\bf Acknowledgments}}\nobreak}
\ack
We would like to thank  
C.~Hofman and H.~Verlinde  for useful discussions. 
JSP is grateful to C.~Vafa for a useful suggestion. 
This work is   supported  by a  Pioneer Fund  from 
the Nederlandse Organisatie voor Wetenschappelijk Onderzoek (NWO).

\vfill\eject

\appendix{A}{The $N=4$ Algebra}

The commutation relations \sac\ for the four supercharges $\bs_\pm$ and $\bbs_\pm$ 
together with the conditions in \hoq\ determine the entire transformation
laws. \def\ep{\epsilon}
In terms of infinitesimal fermionic parameters $\ep_-,\bar\ep_-$
and $\ep_+,\bar\ep_+$, we denote 
$\d = \bar\ep_- \bs_+ +\bar\ep_+\bs_- + \ep_+\bbs_- +\ep_+\bbs_+$.

We start from the consistency nine-plet \ffcccc.
We have
\eqn\jjk{
\eqalign{
\d \phi_{++} &= \bar\ep_+\eta_+ +  \ep_+\bar\eta_+,\cr
\d \phi_{--} &= \bar\ep_-\eta_- +\ep_-\bar\eta_-,\cr
}\qquad
\eqalign{
\d \s &= -\bar\ep_+\eta_- -\ep_-\bar\eta_+,\cr
\d\bar\s &= -\bar\ep_-\eta_+ -\ep_+\bar\eta_-,\cr
}
}
and
\eqn\jjl{
\eqalign{
\d\eta_{+}
     &=+i\ep_+H_0
	+\Fr{1}{2}\ep_+[\s,\bar\s] 
	+\Fr{1}{2}\ep_+[\phi_{++},\phi_{--}]
	+\ep_-[\phi_{++},\bar\s]
,\cr
\d\bar\eta_{+}
     &= -i\bar\ep_+H_0
	-\Fr{1}{2}\bar\ep_+[\s,\bar\s] 
  	+\Fr{1}{2}\bar\ep_+ [\phi_{++},\phi_{--}]
	+\bar\ep_-[\phi_{++},\s]
,\cr
\d\eta_{-} 
	&=+i\ep_-H_0
	-\Fr{1}{2}\ep_-[\s,\bar\s]
	-\Fr{1}{2}\ep_-[\phi_{++},\phi_{--}]
	+\ep_+[\phi_{--},\s]
,\cr
\d\bar\eta_{-} 
 	 &=-i\bar\ep_-H_0
	+\Fr{1}{2}\bar\ep_-[\s,\bar\s] 
  	-\Fr{1}{2}\bar\ep_-[\phi_{++},\phi_{--}]
   	+\bar\ep_+ [\phi_{--},\bar\s]
,\cr
\d H_0
&=
+\Fr{i}{2}\bar\ep_-[\phi_{++},\eta_{-}]
+\Fr{i}{2}\bar\ep_-[\s,\eta_+]
+\Fr{i}{2}\bar\ep_+[\phi_{--},\eta_{+}]
+\Fr{i}{2}\bar\ep_+[\bar\s,\eta_-]
\cr
&\phantom{=}
-\Fr{i}{2}\ep_-[\phi_{++},\bar\eta_-]
-\Fr{i}{2}\ep_-[\bar\s,\bar\eta_+]
-\Fr{i}{2}\ep_+[\phi_{--},\bar\eta_{+}]
-\Fr{i}{2}\ep_+[\s,\bar\eta_-].
}
}

For the holomorphic and anti-holomorphic quartets from $B^{2,0}$ and
$B^{0,2}$, resprectively, we have
\eqn\jjd{
\eqalign{
\d B^{2,0}=& \bar\ep_-\chi_+^{2,0} -\bar\ep_+\chi_-^{2,0},
\cr
\d B^{0,2} 
       =&\ep_-\bar\chi_{+}^{0,2}
	-\ep_+\bar\chi_{-}^{0,2},
\cr
\d \chi_+^{2,0} =& \bar\ep_+ H^{2,0} 
                      -\ep_+[\s, B^{2,0}]
		-\ep_- [\phi_{++}, B^{2,0}], 
\cr
\d \chi_-^{2,0} = &
                    \bar\ep_- H^{2,0}
                  +\ep_-[\bar\s, B^{2,0}]
		+\ep_+ [\phi_{--}, B^{2,0}],
\cr
\d\bar\chi_{+}^{0,2}
     	=& \ep_+ H^{0,2}	
	-\bar\ep_+[\bar\s,B^{0,2}]
	-\bar\ep_-[\phi_{++},B^{0,2}],
\cr
\d\bar\chi_{-}^{0,2} 
	=&\ep_- H^{0,2}
  	+\bar\ep_-[\s, B^{0,2}]
 	+\bar\ep_+[\phi_{--},B^{0,2}],
\cr
\d H^{2,0} = &
	 -\ep_- [\phi_{++},\chi_-^{2,0}]  
	+\ep_-[\eta_+, B^{2,0}] 
	-\ep_-[\bar\s,\chi_+^{2,0}]
	\cr
	&
	-\ep_+ [\phi_{--} ,\chi_+^{2,0} ] 
	-\ep_+[\eta_-,B^{2,0}] 
	-\ep_+[\s,\chi_-^{2,0}],
\cr
\d  H^{0,2} = &
 -\bar\ep_- [\phi_{++},\bar\chi_- ^{0,2}]  
	+\bar\ep_-[\bar\eta_+,  B^{0,2} ] 
	-\bar\ep_-[\s,\bar\chi_+^{0,2} ]
	\cr
	&
	-\bar\ep_+ [\phi_{--}, \bar\chi_+^{0,2} ]  
	-\bar\ep_+[\bar\eta_-, B^{0,2} ] 
	-\bar\ep_+[\bar\s,\bar\chi_-^{0,2} ].
}
}

For the holomorphic and anti-holomorphic quartets from
the connection one-form $A$ we have
\eqn\jjc{
\eqalign{
 \d A^{1,0} 
	=& \ep_+\bar\p^{1,0}_-+\ep_-\bar\p^{1,0}_+ ,
\cr
 \d A^{0,1} 
	=& \bar\ep_+\p^{0,1}_- +\bar\ep_-\p^{0,1}_+ ,
\cr
\d\bar\p_+^{1,0} =&
	-\ep_+ H^{1,0}
	+\bar\ep_+\Dp\bar\s
	+\bar\ep_-\Dp \phi_{++},
\cr
\d\bar\p^{1,0}_- =&
	+\ep_- H^{1,0}
	+\bar\ep_-\Dp\s
	+\bar\ep_+\Dp\phi_{--},
\cr
\d\p_+^{0,1} =&
	-\bar\ep_+ H^{0,1}
	+\ep_+\Dpp\s
	+\ep_-\Dpp\phi_{++},
\cr
\d\p^{0,1}_- =&
	+\bar\ep_- H^{0,1}
	+\ep_-\Dpp\bar\s
	+\ep_+\Dpp\phi_{--},
\cr
\d H^{1,0} =&
	 -\bar\ep_-[\phi_{++},\bar\p^{1,0}_-] 
	+\bar\ep_-[\s,\bar\p^{1,0}_+] 
	-\bar\ep_-\Dp\bar\eta_+
\cr
&	+\bar\ep_+[\phi_{--},\bar\p_+^{1,0}]
	 -\bar\ep_+[\bar\s,\bar\p^{1,0}_-] 
	+\bar\ep_+\Dp\bar\eta_-
\cr
\d H^{0,1} =&
	 -\ep_-[\phi_{++},\p^{0,1}_-] 
	+\ep_-[\bar\s,\p^{0,1}_+] 
	-\ep_-\Dpp\eta_+
\cr
&	+\ep_+[\phi_{--},\p_+^{0,1}]
	 -\ep_+[\s,\p^{0,1}_-] 
	+\ep_+\Dpp\eta_-.
}
}

\vfill\eject
\listrefs
\bye